\documentclass{aa}  
\bibpunct{(}{)}{;}{a}{}{,}
\usepackage{graphicx}
\usepackage{txfonts}
\usepackage{lscape}
\usepackage{ulem}
\usepackage[draft=true]{hyperref}
\newcommand{\degree}{\ensuremath{^\circ}}

\newcommand{\pq}{\ensuremath{P_Q}}
\newcommand{\pu}{\ensuremath{P_U}}

\begin{document} 

\title{The cloudbow of planet Earth observed in polarisation}

\author{ Michael F. Sterzik\inst{1}
  \and
  Stefano Bagnulo\inst{2}
 \and
 Claudia Emde\inst{3}
  \and
 Mihail Manev\inst{3}
 }

\institute{
 European Southern Observatory, Karl-Schwarzschild-Str. 2, D-85748 Garching, Germany 
 \email{msterzik@eso.org}
  \and
  Armagh Observatory and Planetarium, College Hill, Armagh BT61 9DG, UK
\and
Meteorological Institute, Ludwig-Maximilians-University,
  Theresienstr. 37, D-80333 Munich, Germany
}
\date{Received 27-April-2020; Accepted 25-May-2020}

\abstract
{Scattering processes in the atmospheres of planets cause characteristic features that can be particularly well observed in polarisation. 
For planet Earth, both molecular scattering (Rayleigh) and scattering by small particles (Mie) imprint specific signatures in its phase curve.
Polarised phase curves allow us to infer 
physical and chemical properties of the atmosphere like the composition of the gaseous and liquid components, droplet sizes, and refraction indices.}
{An unequivocal prediction of a liquid-water-loaded atmosphere is the existence of a rainbow feature at a scattering angle of around 138-144\degree . Earthshine allows us to observe the primary rainbow in linear polarisation.}
{We observed polarisation spectra of Earthshine using FORS2 at the Very Large Telescope for phase angles from 33\degree\ to 65\degree\ (Sun--Earth--Moon angle). 
The spectra were used to derive the degree of polarisation in the $B$, $V$, $R$, and $I$ passbands and  the phase curve from 33\degree\ to 136\degree\ .
The new observations extend to the smallest phases that can be observed from the ground.}
{The degree of polarisation of planet Earth is increasing for decreasing phase angles downwards of 45\degree\ in the $B$, $V$, $R$, and $I$ passbands.
From comparison of the phase curve observed with models of an Earth-type atmosphere we are able to determine the refractive index of water and to constrain the mean water droplet sizes to $6-7\mu$m. Furthermore, we can retrieve the mean cloud fraction of liquid water clouds to 0.3, and the mean optical depth of the water clouds to values between 10 and 20.}
{Our observations allow us to discern two fundamentally different 
scattering mechanisms of the atmosphere of planet Earth: molecular and particle scattering. 
The physical and chemical properties can be retrieved with high fidelity through suitable inversion of the phase curve. 
Observations of polarimetric phase curves of planets beyond the Solar System shall be extremely valuable for a thorough characterisation of their atmospheres.} 



\maketitle
%
\section{Introduction}

The presence of liquid water is key for a planet to be habitable \citep{Kasting:2003bp}, and the remote detection of water on exoplanets will be a rosetta stone in the search for biosignatures and life on other worlds.   

On present-day Earth, the hydrosphere contains about 2.3\textperthousand\ of Earth's total mass. Thereof, 97\% resides as oceans in the liquid phase, while $\approx$2\% is in the solid phase as ice in glaciers. Only a mass fraction of about $10^{-5}$ is dispersed in the atmosphere as clouds and water vapour. Remote sensing techniques applied to Earth observations take advantage of specific signatures that the liquid water phase imprints on light by reflection on ocean surfaces or by cloud scattering. 

Observations of polarisation as a function of the phase angle (or scattering angle) allow a detailed diagnosis of the physical and chemical properties of liquid droplets in a planet's atmosphere, even for small quantities. Spherical particles produce a strongly polarised peak known as `rainbow scattering' caused by refraction and internal reflection of light in droplets. The rainbow scattering angle $\beta$ is about 139\degree\ for water, and its contrast is much higher in polarisation than in flux.  The mechanism is commonly described by Mie scattering and mainly depends on particle size distribution and index of refraction. Rainbows seen from above are often dubbed `cloudbows' and may appear less colourful because of the smaller particle sizes involved.

In the Solar System, the polarised phase curve of planet Venus provides an opportunity to reliably retrieve its atmospheric main constituents (in this case a concentrated solution of sulfuric acid) and sizes (around $1\mu$m) from ground-based observations, and demonstrates the power of measuring and interpreting phase angles around the rainbow scattering angle  \citep{1974JAtS...31.1137H}.

Evidently, this physical mechanism and its detectability by  polarisation has applications beyond the Solar System. 
\citet{2007AsBio...7..320B} explored the effects of rainbow polarisation in the search for habitable planets. He estimated a disc-integrated rainbow peak between 12\% and 15\% (fractional polarisation) for Earth if seen as an exoplanet. In order to prepare the quest to find and interpret observations of a `second Earth', radiative transfer models of Earth-type model planets with clouds have been calculated that include proper treatment of scattering and polarisation. 
\citet{Stam:2008ij} simulated polarisation spectra over the entire phase curve in order to investigate the complex interaction of light with different surface types underneath an atmosphere which itself hosts clouds. This work was extended by using more realistic and patchy clouds over inhomogeneous surfaces in  \citet{Karalidi:2012kx}. Calculations in \citet{Karalidi:2012fc} corroborated the existence of a robust rainbow signature for Earth-like planets, even if clouds are composed from mixtures of liquid droplets and ice crystals.   

However, whether or not the rainbow is visible to putative distant observers of planet Earth is unclear. It is also unclear whether or not these observers would be able to derive chemical and physical properties of its atmosphere by remote sensing, as we can do for Venus. 

So far, polarisation measurements of the rainbow of Earth from above have only been performed by airborne instruments \citep{Alexandrov:2012bd} or satellites. Using spaceborne polarisation measurements with POLDER onboard the PARASOL satellite, \citet{Breon:1998tv} provided quantitative polarisation maps of Earth in polarisation. Single scattering by cloud droplets located at the cloud top cause multiple cloudbow features. The inversion of the observed polarised reflectance maps in three spectral bands (0.44, 0.67 and 0.86 $\mu$m) resulted in narrow droplet distributions ($\varv_{\rm eff}\approx 0.02$) and droplet effective radii between 8 and 10 $\mu$m. 
Knowledge of the effective sizes, shapes, and mass concentrations of cloud droplets is highly relevant for the energy budget in the atmosphere of Earth. Droplet sizes relate to cloud albedos, distinguish clean and polluted clouds \citep{Peng:2002gy}, and represent a sensitive input parameter for global climate simulations; see e.g. \citet{Gultepe:2004kh}. 

Furthermore, the spatial and temporal distribution of the thermodynamical phases of clouds and the contributions of mixed-phase and ice clouds for climate sensitivity in the global climate model are highly relevant, but unfortunately not well constrained \citep{Lohmann:2018ke}.  In particular, disc-integrated whole-Earth observations are scarce and hard to obtain.
\citet{Goloub:2000ve} inferred the presence of an ice phase in the upper cloud layers from POLDER measurements.
\citet{2019RAA....19..117W} used a specific set of PARASOL data to re-construct the diurnal variation of disc-integrated polarisation in its three spectral bands at a phase angle of 55\degree , but these do not cover the rainbow angle. Their values are generally consistent with expectations from simple models, but are very limited in terms of viewing geometry and therefore do not allow further constraint of cloud or surface properties.

In principle, disc-integrated flux and polarisation information from the whole Earth can be obtained with Earthshine measurements. 
Earthshine is sunlight scattered by Earth and indirectly reflected from the lunar surface back to Earth, where it can be
observed from the ground. 
This mimics observations of Earth as an exoplanet. Earthshine polarisation observations have been pioneered by \citet{1957SAnAp...4....3D}, who determined Earth's fractional polarisation for several phase angles sampled between 30\degree\ and 140\degree. While his measurements nicely describe a Rayleigh scattering atmosphere with the expected peak polarisation around quadrature, it is interesting to note that no rainbow feature \sout{at small phase-angles} is present, likely because of insufficient phase sampling.  

Modern polarisation measurements of Earthshine allow us to infer new and more detailed insight into Earth as seen as an exoplanet. Using spectropolarimetry in optical wavelengths, \citet{Sterzik:2012gk} derived fractional contributions of clouds and ocean surfaces and was able to distinguish visible areas of vegetation as small as 10\% in two different aspect geometries of Earth, both close to quadrature. \citet{2013PASJ...65...38T}, \citet{Bazzon:2013gl}, and \citet[hereafter Paper 1]{Sterzik:2019fh} 
extended the phase coverage of Earthshine observations to phase angles between 30\degree\ and 140\degree. All three works are - within the expected variations caused by different sceneries and cloud coverage -- consistent with each other and also agree broadly with the models, except at longer wavelengths, where the observed polarisation was always higher than model predictions.

The impact of ice clouds and the potential visibility of the ocean glint on the polarisation spectra of Earth was recently explained in more detail through 3D vector radiative transfer calculations by \citet{Emde:2017eea}. Cloud composition, stratification, and patchyness all have significant and observable impact on the strength and spectral shape of Earth's polarisation spectra, and must be properly included in the simulations. 
Direct reflection of sunlight either directly on the surface of the ocean, or on cloud decks, also contributes to an enhanced polarisation, and can be expected ubiquitously \citep{Trees:2019fc}.  These simulations of realistic Earth models included the treatment of mixtures of liquid and ice droplets, and predict the existence of a robust cloudbow feature in the polarisation phase curve for adequate choices of size parameters. A framework for simulations of comprehensive models of Earth's disc-integrated Stokes vectors has also been put forward by \citet{2015IJAsB..14..379G}.  

However, until now, no indication of a cloudbow feature has been observed in integrated whole-disc observations of Earth.
Here, we report a clear detection of the cloudbow through polarisation observations of Earthshine for the first time. 
The measurements allow a relatively robust retrieval of the refractory index and  characteristic sizes of the scattering agent, in this case liquid water. 
The detection of enhanced polarisation caused by cloudbows on other planets may become a realistic possibility and would contribute to characterisation of the composition of exo-atmospheres and help to constrain bio-markers in other worlds. 

\section{Observations}\label{Sect_Observations}

We used the FORS2 instrument mounted at the Cassegrain focus of the Antu telescope of the VLT during four suitable monthly observing time windows closely following the new moon from late October 2019 to late January 2020.
The FORS2 instrument with its wollaston prism and rotating retarder waveplate enables direct measurement of the (wavelength-dependent) quantities $\pq=Q/I$ and $\pu=U/I$, and thus the fractional polarisation $P(\lambda)$ (or degree of polarisation, referred to in this work as simply `polarisation', as a percentage) defined as 
\begin{equation}
P = \sqrt{(\pq^2 + \pu^2)}.
\label{eq:pol}
\end{equation} 
The angle of polarisation ($\phi$) may be obtained from
\begin{equation}
\tan( 2 \cdot \phi) = {U/Q}.
\label{eq:ang} 
\end{equation}
Observations were obtained with grism 300V, which spans the optical range between 3600\;\AA\; and 9200\;\AA, resulting in a spectral resolution of $\approx$220 with a 2\arcsec\ slit width. The slit was always oriented to cross the eastern part of the lunar limb. Half of the available detector therefore contains the Earthshine signal on the lunar surface, while the other half contains empty sky. This allows to linearly extrapolate and substract the sky background that otherwise contaminates Earthshine, as explained in \citet{Hamdani:2006cn}.   

The observation dates were chosen to be able to observe Earthshine at phase angles  $\alpha$ of the Sun--Earth--Moon (S--E--M) geometry that allow sampling of the suspected cloudbow feature around $\alpha \approx 40\degree$ ($\alpha$ relates to the scattering angle $\beta$ as $\alpha = 180\degree - \beta$). Observations in this geometrical configuration are challenging, and are 
only possible during a short interval of time of only one evening twilight per lunar cycle.
Observations shortly after evening twilight in Chile sample large portions of the illuminated Pacific ocean.  We chose four observing epochs, listed in Table~\ref{Tab:Log}. To maintain consistency with the labels used in paper 1, we denote the new datasets as H.x, I.x, J.x and K.x, respectively. The dates of observations cover phase angles $\alpha$ of 33, 35, 37, 42, 44, 50, 53, 54, 55, 56, 57, 62, 63 and 65 degrees. As we show below, these phases allow us to sample the cloudbow feature roughly around its maximum expected polarisation, albeit on different observing dates. We did not have access to even smaller phase angles for which cloudbow polarisation is expected to cease. 

In order to largely suppress systematic uncertainties of the calibration of dual-beam retarder polarimeters, we used the so-called `beam swapping technique', and observed the target (the lunar limb) with the retarder waveplate orientation consecutively set to position angles 0, 22.5, 45, ..., 337.5\degree . The number of distinct retarder settings is indicated in Table~\ref{Tab:Log} by the parameter $N_{\rm cyc}$ and is usually equal to 16.  However, occasionally, in particular at the lowest phase angles, observing time was not sufficient to conclude a full cycle of 16 retarder settings. In these cases, only eight or four angles of distinct retarder settings were performed. Albeit with lower signal-to-noise ratio, this is still sufficient to suppress most calibration errors, and `null polarisation' spectral profiles are always healthy with statistical errors around zero \citep[for their definition see][]{2009PASP..121..993B}.

The spectra (and therefore the derived values for the fractional polarisation $P$) for the observations at a phase angle of $\alpha = 33$\degree\ (dataset "K.1") and $\alpha = 35$\degree\ (dataset "J.1") are the most uncertain compared to all other spectra. For these spectra, the total time observing Earthshine during twilight was small, and in particular dataset J.1 may suffer from additional systematic errors due to the acquisition of only two retarder positions to derive the Stokes parameters. In principle, two distinct retarder positions are sufficient to determine the two Stokes parameters $Q$ and $U$ independently, but any systematic error introduced during the observation may then not be cancelled out by the beam-swapping techniques. We estimate these errors by subdividing a different observing sequence (J.3, which has $N_{\rm cyc}$=8) into four subsets of two settings only, and determine their polarisation spectra correspondingly. These subcycles are indicated by J.3-1 to J.3-4 in Table~\ref{Tab:Log}. Typical errors of about 0.5\% are introduced in the polarisation, but these errors appear  still  lower than those introduced by the uncertainties on the lunar depolarisation. These have an effect of 1-2\%, as can be seen in Table~\ref{Tab:PEarth}. We therefore consider the J.1 dataset in the analysis with its higher systematic errors.
Although both spectra J.1 and K.1 have comparatively low S/N, we include them in the subsequent analysis (but take into account their higher statistical errors).    

We refer to Paper~1 for a full description of the data-reduction techniques, method of background subtraction, and correction for lunar depolarisation. The last two steps are crucial for a proper analysis. Due to a viewing geometry close to the new moon, the contamination with moonshine background is even lower than in the datasets presented in Paper~1. Linear background extrapolation and subtraction works well and gives reliable results. Here we also consider data from Paper~1 that refer to the "Pacific" viewing geometry, and therefore we reprocessed all datasets consistently with minor improvements. We report all values (and their errors) including updated values for datasets B.x, E.x, and G.x in Tables~\ref{Tab:Log} and \ref{Tab:PEarth}. Differences between the old and new data reductions are typically below 1\% for Earthshine polarisation, and always within the systematic errors for Earth polarisation.

\section{Results}

\subsection{Polarisation spectra}

\begin{figure*}[t]
\resizebox{\hsize}{!}{\includegraphics[width=\textwidth, trim=50 50 50 50]{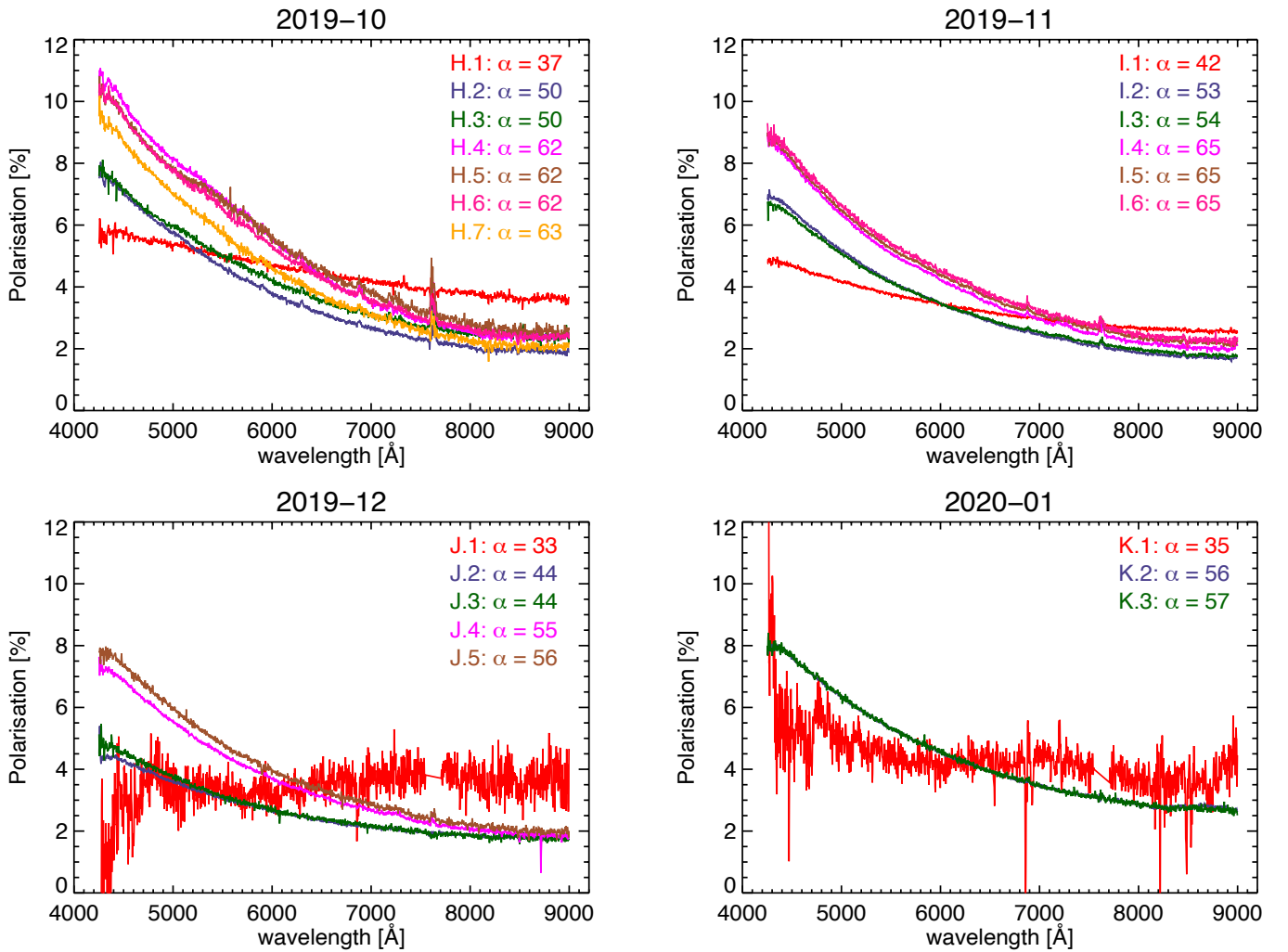}} 
\caption{Spectra of the fractional polarisation of Earthshine obtained with FORS2 in the 2019-10 to 2020-01
observing campaigns. Individual observations are identified  by colours specified in the legend, where each spectrum is denoted by its observation ID and its actual phase angle $\alpha$ in Table~\ref{Tab:Log}.}
\label{Fig:AllSpectra}
\end{figure*}

Earthshine polarisation spectra are shown in Fig.~\ref{Fig:AllSpectra} for our four new observing epochs. Spectra are similar and tend to decrease with increasing wavelength and to increase for increasing phase-angle in the blue spectral range; they exhibit similar polarisation values in the red part of the spectrum, independent of phase-angle. Lower phase angles (37\degree\ and 42\degree ) show polarisation that is only weakly dependent on the wavelength in contrast to higher phase angles. The polarisation spectra for $\alpha$=33\degree\ (J.1) and 35\degree\ (K.1) are spectrally rather flat, albeit noisier due to low {\it S/N}; because of a very low flux signal of Earthshine as compared to twilight sky background, their spectra are particularly uncertain around both ends of the available wavelength coverage, and in the region of the O$_2$-A line around 7700~\AA\ (which is omitted in Fig.~\ref{Fig:AllSpectra} for these two cases).

All spectra have been observed with the grism 300V, and have a spectral resolution of approximately 220 (which corresponds to a resolving power of $\approx$ 30 \AA), given a slit width of 2\arcsec.  In order to better analyse the spectra, we derived the degree of polarisation of Earthshine $P^{\rm ES}$ for the four bandpasses  $B, V, R$ and $I$, averaged over the wavelengths $P_B$: 4350 -- 4550~\AA,  $P_V$: 5450 -- 5650~\AA, $P_R$: 6450 -- 6650~\AA, and $P_I$: 8050 -- 8650~\AA. The simultaneous measurement of Stokes $Q$ and $U$ also allows us to calculate the angle of polarisation $\phi$. This can be compared to the angle $\Phi$ between the normal of the scattering plane and the celestial north pole. As shown in Paper 1, we find very good agreement between these values, which are relatively independent of wavelength. For cases J.1 and K.1, the differences are larger and of the order of 10\degree . This may be due to residuals of twilight sky that might not have been fully removed and contaminates the spectra. 


\subsection{Polarisation phase curves}\label{Phasecurve}

\begin{figure}[t]
\resizebox{\hsize}{!}{\includegraphics[trim=50 50 50 50]{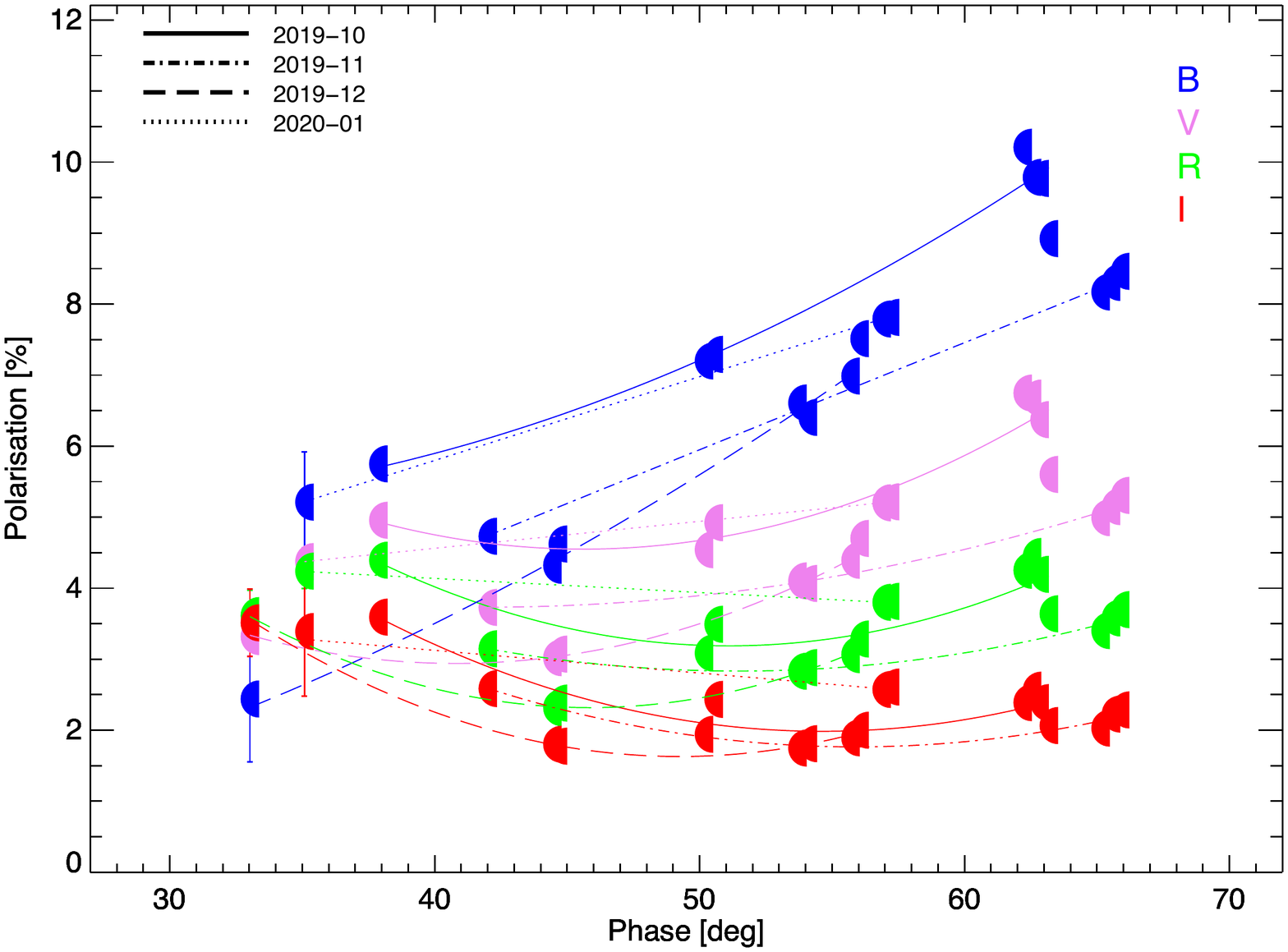}} 
\caption{Fractional polarisation of Earthshine $P^{\rm ES}$ as a function of the phase angle (Sun -- Earth -- Moon). The bandpasses correspond 
approximately to traditional $B$, $V$, $R,$ and $I$ bands (see text). Lines connect observation cycles that belong to sequences of observations during consecutive nights, while different line styles differentiate observations separated by months (see Table~\ref{Tab:Log}).}
\label{Fig:AllPA_own}
\end{figure}

The fractional polarisation of Earthshine $P^{\rm ES}$ extracted from the spectra for the four bandpasses (as listed in Table~\ref{Tab:Log}) is plotted as a function of phase angle $\alpha$ in Fig.~\ref{Fig:AllPA_own}. Different colours indicate different bands $B$, $V$, $R$ and $I$.  Symbol sizes are always larger than the statistical errors calculated from the spectra, except for datasets J.1 and K.1.
For phase angles larger than 45\degree,
polarisation $P$ is always higher in blue than red, and is increasing for increasing $\alpha$ in all bands. Even for a narrow range of phase angles of a few degrees there is significant variation of $P$ (up to 2\% in the blue and up to 1\% in the red) when observed at different epochs. All these findings are qualitatively consistent with observations at higher phase angles reported in Paper~1. 

$P^{\rm ES}$ reaches minimum values in all bands for $\alpha$ around 45\degree, $\approx$5\% in $B$ and $\approx$2\% in $R$. For further decreasing $\alpha$, $P$ increases again for all bands. Observations between $\alpha \approx 35$\degree\ and 37\degree\ appear to sample a local peak, which reaches $P^{\rm ES}_B$=6\% in $B$, and around 3-4\% in $R$.  Observations at $\alpha \approx 33$\degree\ are distinctively lower, between 2-4\% for all bands; the (uncertain) value in $B$ is lowest.  

The values of polarisation corrected for lunar depolarisation $P^{\rm E}$ in the four bands are listed in Table~\ref{Tab:PEarth}. Two types of errors are given in the table: statistical and systematic. Statistical errors (in brackets) are calculated from null profiles.  The systematic errors of $P^{\rm E}$ come from uncertainties in determining the lunar albedo at the observing slit position. As in Paper 1, the relation of a wavelength-dependant polarisation efficiency as a function of the lunar albedo is assumed \citep{Bazzon:2013gl}. We calculate the combined errors of $P^{\rm E}$ as the root of  squares of the sum of statistical and systematic errors. 

\begin{figure}[t] 
\resizebox{\hsize}{!}{\includegraphics[trim=50 50 50 50]{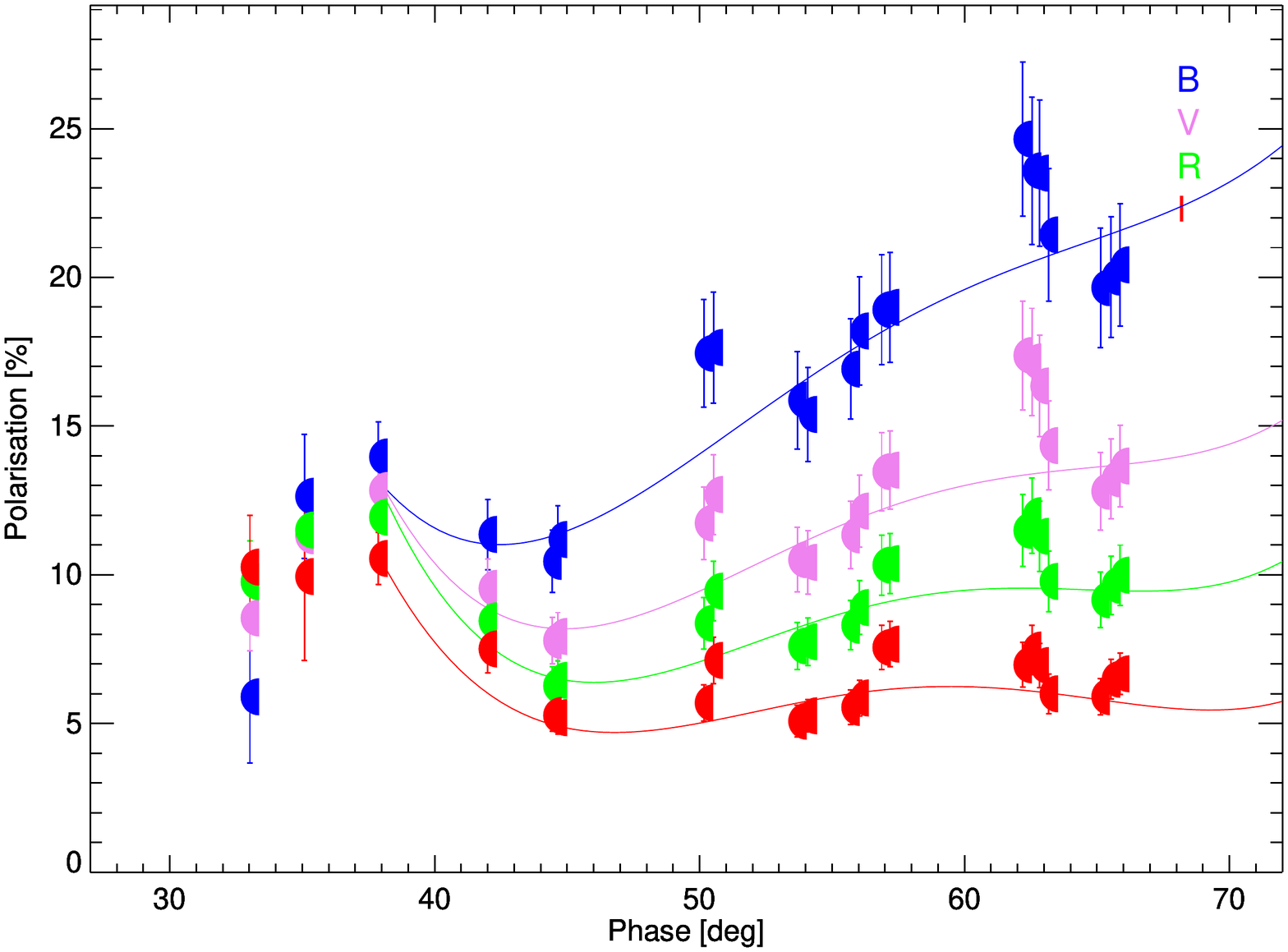}} 
\caption{Fractional polarisation of Earth $P^{\rm E}$  after correction for lunar depolarisation effects as a function of the phase angle  (Sun--Earth--Moon).
Colours indicate the values for different bandpasses $B$, $V$, $R,$ and $I$. The lines fit a low-order polynomial to the data in each band, and help to guide the eye to follow the general trends of the phase curves between 35\degree\ and 70\degree . }
\label{Fig:Earthpol}
\end{figure}

$P^{\rm E}$ and their errors are plotted in Fig.~\ref{Fig:Earthpol} as a function of the phase angle $\alpha$. While the lunar depolarisation correction mainly re-scales the fractional polarisation $P^{\rm ES}$ by a factor of $\approx5/2$, it also introduces a slight modification of the spectral shape due to its weak wavelength dependence. 

The general shape of the phase curves is similar for the polarisation of Earthshine $P^{\rm ES}$ (Fig.~\ref{Fig:AllPA_own}) and Earth $P^{\rm E}$ (Fig.~\ref{Fig:Earthpol}) in all four bandpasses. In order to help to guide the eye to follow the general trends of the phase curve in each band, the full lines in Fig.~\ref{Fig:Earthpol} show a low-order polynomial fit to the data in each bandpass. The local minimum around $\alpha\approx 45$\degree , the increase towards the local maximum at $\alpha\approx 35$\degree\ , and the decline for  $\alpha\approx 33$\degree\ are evident.

Following the spectral shape, the fractional polarisation for $\alpha \ge  35$\degree\ is always ordered according to $P^{\rm E}_B > P^{\rm E}_V> P^{\rm E}_R > P^{\rm E}_I$. Interestingly,  for $\alpha = 33$\degree\ this order is reversed, and polarisation slightly increases with increasing wavelength: $P^{\rm E}_I \sim P^{\rm E}_R \sim P^{\rm E}_V  > P^{\rm E}_B$. 

\setcounter{table}{1}
\begin{table*}
\caption{Degree of polarisation $P^{\rm E}$ in our bandpasses corrected for lunar depolarisation. The determination of rotation angle $\alpha_{\rm Moon}$ of the moon with respect to the acquisition images and mean albedo $\overline{a}_{603}$ for each observed region are explained in the text. Statistical errors of $P^{\rm E}$ (in brackets) are calculated from the null-profiles. Systematic errors of $P^{\rm E}$ are derived from the systematic errors to determine the lunar albedo $\overline{a}_{603}$. }
\label{Tab:PEarth} 
\centering
\begin{tabular}{lrlrrrrr}
\hline\hline
ID & $\alpha_{\rm Moon}$ & $\overline{a}_{603}$ & $P_B^{{\rm E}}$[\%] & $P_V^{{\rm E}}$[\%] & $P_R^{{\rm E}}$[\%] &$ P_I^{{\rm E}}$[\%]  \\ 
\hline 
B.1 & 140.0 & 0.176(0.007) & 34.4(0.1)$^{+ 1.6}_{- 1.7}$ & 22.7(0.1)$^{+ 1.1}_{- 1.1}$ & 19.1(0.1)$^{+ 0.9}_{- 0.9}$ & 16.8(0.4)$^{+ 0.8}_{- 0.8}$ & \\
B.2 & 138.5 & 0.177(0.006) & 40.3(0.2)$^{+ 1.8}_{- 1.8}$ & 26.3(0.1)$^{+ 1.2}_{- 1.2}$ & 21.7(0.1)$^{+ 1.0}_{- 1.0}$ & 16.9(0.3)$^{+ 0.7}_{- 0.8}$ & \\
B.3 & 138.5 & 0.177(0.006) & 39.6(0.1)$^{+ 1.6}_{- 1.7}$ & 26.7(0.1)$^{+ 1.1}_{- 1.1}$ & 22.0(0.1)$^{+ 0.9}_{- 0.9}$ & 17.8(0.4)$^{+ 0.7}_{- 0.8}$ & \\
B.4 & 140.5 & 0.179(0.007) & 36.1(0.1)$^{+ 1.7}_{- 1.8}$ & 24.8(0.1)$^{+ 1.2}_{- 1.2}$ & 21.8(0.1)$^{+ 1.0}_{- 1.1}$ & 19.7(0.5)$^{+ 0.9}_{- 1.0}$ & \\
B.5 & 137.5 & 0.179(0.007) & 34.4(0.3)$^{+ 1.6}_{- 1.6}$ & 24.2(0.1)$^{+ 1.1}_{- 1.1}$ & 22.0(0.1)$^{+ 1.0}_{- 1.0}$ & 20.4(0.8)$^{+ 0.9}_{- 1.0}$ & \\[1mm]
E.1 & 110.0 & 0.174(0.007) & 14.6(0.1)$^{+ 0.7}_{- 0.7}$ &  9.7(0.0)$^{+ 0.4}_{- 0.4}$ &  7.0(0.0)$^{+ 0.3}_{- 0.3}$ &  4.8(0.1)$^{+ 0.2}_{- 0.2}$ & \\
E.2 & 113.0 & 0.175(0.007) & 21.2(0.1)$^{+ 1.0}_{- 1.0}$ & 13.7(0.1)$^{+ 0.6}_{- 0.6}$ &  9.5(0.1)$^{+ 0.4}_{- 0.5}$ &  6.0(0.1)$^{+ 0.3}_{- 0.3}$ & \\
E.4 & 118.0 & 0.179(0.007) & 26.6(0.1)$^{+ 1.2}_{- 1.3}$ & 17.5(0.1)$^{+ 0.8}_{- 0.8}$ & 12.2(0.1)$^{+ 0.6}_{- 0.6}$ &  7.4(0.2)$^{+ 0.3}_{- 0.4}$ & \\
E.6 & 118.0 & 0.183(0.007) & 25.9(0.6)$^{+ 1.2}_{- 1.3}$ & 17.3(0.1)$^{+ 0.8}_{- 0.8}$ & 12.4(0.1)$^{+ 0.6}_{- 0.6}$ &  7.9(0.2)$^{+ 0.4}_{- 0.4}$ & \\[1mm]
G.1 & 103.0 & 0.179(0.010) & 37.2(0.5)$^{+ 2.4}_{- 2.5}$ & 30.0(0.5)$^{+ 1.9}_{- 2.0}$ & 22.0(0.7)$^{+ 1.4}_{- 1.5}$ & 18.6(0.7)$^{+ 1.2}_{- 1.2}$ & \\
G.2 &  99.0 & 0.179(0.010) & 38.5(0.2)$^{+ 2.6}_{- 2.7}$ & 29.5(0.2)$^{+ 2.0}_{- 2.1}$ & 22.6(0.2)$^{+ 1.5}_{- 1.6}$ & 19.8(0.3)$^{+ 1.3}_{- 1.4}$ & \\
G.4 &  98.0 & 0.177(0.012) & 33.0(0.4)$^{+ 2.6}_{- 2.7}$ & 24.5(0.2)$^{+ 1.9}_{- 2.0}$ & 21.3(0.1)$^{+ 1.7}_{- 1.8}$ & 18.6(0.3)$^{+ 1.4}_{- 1.5}$ & \\
G.5 &  93.0 & 0.176(0.012) & 31.3(0.3)$^{+ 2.5}_{- 2.6}$ & 23.9(0.3)$^{+ 1.9}_{- 2.0}$ & 17.2(0.3)$^{+ 1.4}_{- 1.4}$ & 14.3(0.7)$^{+ 1.1}_{- 1.2}$ & \\
G.7 &  90.0 & 0.174(0.012) & 13.8(0.1)$^{+ 1.2}_{- 1.2}$ & 11.2(0.1)$^{+ 0.9}_{- 1.0}$ &  9.1(0.1)$^{+ 0.8}_{- 0.8}$ & 10.2(0.2)$^{+ 0.9}_{- 0.9}$ & \\
G.9 &  90.0 & 0.173(0.013) & 13.6(0.1)$^{+ 1.2}_{- 1.3}$ & 11.2(0.1)$^{+ 1.0}_{- 1.1}$ &  8.8(0.1)$^{+ 0.8}_{- 0.9}$ &  8.0(0.6)$^{+ 0.7}_{- 0.8}$ & \\[1mm]
H.1 &  87.0 & 0.171(0.014) & 14.0(0.2)$^{+ 1.4}_{- 1.5}$ & 12.8(0.1)$^{+ 1.3}_{- 1.3}$ & 11.9(0.1)$^{+ 1.2}_{- 1.3}$ & 10.5(0.1)$^{+ 1.0}_{- 1.1}$ & \\
H.2 &  90.0 & 0.171(0.014) & 17.4(0.1)$^{+ 1.7}_{- 1.9}$ & 11.7(0.1)$^{+ 1.2}_{- 1.3}$ &  8.4(0.1)$^{+ 0.8}_{- 0.9}$ &  5.7(0.2)$^{+ 0.6}_{- 0.6}$ & \\
H.3 &  90.0 & 0.170(0.015) & 17.6(0.4)$^{+ 1.8}_{- 1.9}$ & 12.7(0.1)$^{+ 1.3}_{- 1.4}$ &  9.5(0.1)$^{+ 1.0}_{- 1.0}$ &  7.1(0.2)$^{+ 0.7}_{- 0.8}$ & \\
H.4 &  97.0 & 0.170(0.015) & 24.6(0.1)$^{+ 2.5}_{- 2.7}$ & 17.4(0.1)$^{+ 1.8}_{- 1.9}$ & 11.5(0.1)$^{+ 1.2}_{- 1.2}$ &  7.0(0.2)$^{+ 0.7}_{- 0.8}$ & \\
H.5 &  97.0 & 0.169(0.014) & 23.6(0.2)$^{+ 2.4}_{- 2.5}$ & 17.2(0.2)$^{+ 1.7}_{- 1.9}$ & 12.0(0.2)$^{+ 1.2}_{- 1.3}$ &  7.5(0.3)$^{+ 0.8}_{- 0.8}$ & \\
H.6 &  97.0 & 0.169(0.014) & 23.5(0.1)$^{+ 2.4}_{- 2.5}$ & 16.3(0.1)$^{+ 1.6}_{- 1.8}$ & 11.3(0.1)$^{+ 1.1}_{- 1.2}$ &  6.9(0.2)$^{+ 0.7}_{- 0.7}$ & \\
H.7 &  97.0 & 0.168(0.014) & 21.4(0.2)$^{+ 2.2}_{- 2.3}$ & 14.3(0.1)$^{+ 1.4}_{- 1.5}$ &  9.8(0.1)$^{+ 1.0}_{- 1.0}$ &  6.0(0.2)$^{+ 0.6}_{- 0.6}$ & \\[1mm]
I.1 & 103.0 & 0.168(0.014) & 11.4(0.1)$^{+ 1.1}_{- 1.2}$ &  9.5(0.0)$^{+ 1.0}_{- 1.0}$ &  8.5(0.0)$^{+ 0.8}_{- 0.9}$ &  7.5(0.2)$^{+ 0.8}_{- 0.8}$ & \\
I.2 & 106.0 & 0.168(0.014) & 15.9(0.1)$^{+ 1.6}_{- 1.7}$ & 10.5(0.0)$^{+ 1.0}_{- 1.1}$ &  7.6(0.0)$^{+ 0.8}_{- 0.8}$ &  5.1(0.1)$^{+ 0.5}_{- 0.5}$ & \\
I.3 & 109.0 & 0.168(0.014) & 15.4(0.1)$^{+ 1.5}_{- 1.6}$ & 10.4(0.0)$^{+ 1.0}_{- 1.1}$ &  7.7(0.0)$^{+ 0.8}_{- 0.8}$ &  5.3(0.1)$^{+ 0.5}_{- 0.6}$ & \\
I.4 & 110.0 & 0.169(0.014) & 19.7(0.1)$^{+ 1.9}_{- 2.1}$ & 12.8(0.0)$^{+ 1.3}_{- 1.4}$ &  9.2(0.1)$^{+ 0.9}_{- 1.0}$ &  5.9(0.1)$^{+ 0.6}_{- 0.6}$ & \\
I.5 & 112.0 & 0.169(0.014) & 20.0(0.1)$^{+ 2.0}_{- 2.1}$ & 13.2(0.1)$^{+ 1.3}_{- 1.4}$ &  9.7(0.1)$^{+ 0.9}_{- 1.0}$ &  6.5(0.1)$^{+ 0.6}_{- 0.7}$ & \\
I.6 & 115.0 & 0.170(0.014) & 20.4(0.1)$^{+ 2.0}_{- 2.1}$ & 13.7(0.1)$^{+ 1.3}_{- 1.4}$ & 10.0(0.1)$^{+ 1.0}_{- 1.0}$ &  6.7(0.2)$^{+ 0.6}_{- 0.7}$ & \\[1mm]
J.1 & 107.0 & 0.170(0.014) &  5.9(2.2)$^{+ 0.6}_{- 0.6}$ &  8.5(0.7)$^{+ 0.8}_{- 0.9}$ &  9.8(1.0)$^{+ 0.9}_{- 1.0}$ & 10.3(1.4)$^{+ 1.0}_{- 1.1}$ & \\
J.2 & 111.0 & 0.170(0.014) & 10.5(0.1)$^{+ 1.0}_{- 1.1}$ &  7.8(0.0)$^{+ 0.8}_{- 0.8}$ &  6.3(0.0)$^{+ 0.6}_{- 0.7}$ &  5.3(0.1)$^{+ 0.5}_{- 0.5}$ & \\
J.3 & 111.0 & 0.170(0.014) & 11.2(0.2)$^{+ 1.1}_{- 1.2}$ &  7.9(0.1)$^{+ 0.8}_{- 0.8}$ &  6.5(0.1)$^{+ 0.6}_{- 0.7}$ &  5.2(0.2)$^{+ 0.5}_{- 0.5}$ & \\
\textsl{  J.3-1} & 111.0 & 0.170(0.014) & 10.6(0.1)$^{+ 1.0}_{- 1.1}$ &  8.0(0.2)$^{+ 0.8}_{- 0.8}$ &  7.1(0.1)$^{+ 0.7}_{- 0.7}$ &  6.7(0.1)$^{+ 0.6}_{- 0.7}$ & \\
\textsl{  J.3-2} & 111.0 & 0.170(0.014) & 12.4(0.2)$^{+ 1.2}_{- 1.3}$ &  9.8(0.1)$^{+ 0.9}_{- 1.0}$ &  7.6(0.1)$^{+ 0.7}_{- 0.8}$ &  6.6(0.1)$^{+ 0.6}_{- 0.7}$ & \\
\textsl{  J.3-3} & 111.0 & 0.170(0.014) & 12.5(0.1)$^{+ 1.2}_{- 1.3}$ &  9.9(0.1)$^{+ 1.0}_{- 1.0}$ &  7.7(0.1)$^{+ 0.7}_{- 0.8}$ &  6.8(0.1)$^{+ 0.7}_{- 0.7}$ & \\
\textsl{  J.3-4} & 111.0 & 0.170(0.014) & 10.5(0.2)$^{+ 1.0}_{- 1.1}$ &  7.9(0.1)$^{+ 0.8}_{- 0.8}$ &  7.0(0.1)$^{+ 0.7}_{- 0.7}$ &  6.7(0.1)$^{+ 0.7}_{- 0.7}$ & \\
J.4 & 109.0 & 0.171(0.014) & 16.9(0.1)$^{+ 1.6}_{- 1.7}$ & 11.3(0.1)$^{+ 1.1}_{- 1.2}$ &  8.3(0.1)$^{+ 0.8}_{- 0.9}$ &  5.5(0.2)$^{+ 0.5}_{- 0.6}$ & \\
J.5 & 109.0 & 0.171(0.014) & 18.2(0.3)$^{+ 1.7}_{- 1.9}$ & 12.1(0.2)$^{+ 1.2}_{- 1.2}$ &  8.9(0.2)$^{+ 0.9}_{- 0.9}$ &  5.9(0.2)$^{+ 0.6}_{- 0.6}$ & \\[1mm]
K.1 & 112.0 & 0.171(0.014) & 12.6(1.7)$^{+ 1.2}_{- 1.3}$ & 11.3(0.5)$^{+ 1.1}_{- 1.1}$ & 11.5(0.7)$^{+ 1.1}_{- 1.2}$ &  9.9(2.6)$^{+ 0.9}_{- 1.0}$ & \\
K.2 & 112.0 & 0.171(0.014) & 18.9(0.1)$^{+ 1.8}_{- 1.9}$ & 13.5(0.1)$^{+ 1.3}_{- 1.4}$ & 10.3(0.0)$^{+ 1.0}_{- 1.0}$ &  7.6(0.1)$^{+ 0.7}_{- 0.8}$ & \\
K.3 & 112.0 & 0.172(0.014) & 19.0(0.3)$^{+ 1.8}_{- 1.9}$ & 13.5(0.2)$^{+ 1.3}_{- 1.3}$ & 10.4(0.1)$^{+ 1.0}_{- 1.0}$ &  7.7(0.2)$^{+ 0.7}_{- 0.8}$ & \\[1mm]
\hline
\hline
\end{tabular}
\end{table*}

\section{Comparison with models}

The local minimum of fractional polarisation in the phase curve of Earth observed with high accuracy around 45\degree\ and the local maximum around 35\degree-37\degree\ are characteristic for a primary rainbow. 
Their quantitative explanation requires multiple scattering radiative-transfer simulations, which include optical properties of cloud droplets calculated following Mie theory, as for example employed for the interpretation of the polarisation phase curve of Venus \citep{1974JAtS...31.1137H}. 
In the following, we use increasingly complex models to study if and how the observed phase curves allow us to retrieve the main physical and chemical properties of the atmosphere observed.   

\subsection{Heuristic model}\label{heuristic}

First, we introduce a simple model with two physical components.
For the Rayleigh component, we assume a three parameter model following \citet{2005SoSyR..39...45K}, which was already applied to fit the phase curve in Paper 1; see \citet{Sterzik:2019fh}: 
\begin{equation}\label{eq:rayleigh}
P_{\rm Ray}(\alpha,\lambda) = \frac{(\sin^2 (\alpha - \Delta\alpha(\lambda))^{W(\lambda)}}{1 + \cos^2 (\alpha - \Delta\alpha(\lambda)) + dePol(\lambda)}
.\end{equation}
The three parameters $\Delta\alpha$, $W$, and $dePol$ of Eq.~\ref{eq:rayleigh} parameterise a possible phase shift, skewness, and depolarisation to be accounted for in realistic atmospheres. All three parameters may depend on wavelength, and may therefore be different for each observing bandpass. The parameter $dePol$, in particular, reflects which maximum polarisation can be attained close to quadrature in the presence of a non-zero surface albedo. Multiple scattering, for example in opaque clouds, also tends to increase $dePol$ and reduce the fractional polarisation at all phase angles. 

Rayleigh polarisation in Eq.~\ref{eq:rayleigh} is small and approaches zero for 
both low and high phase angles. On the other hand, a Mie component adds to total polarisation specifically at low phase angles, while it is negligible at high phase angles. We may therefore add-up both components independently,  and approximate the total polarisation using relation Eq.~\ref{eq:rayleighmie}: 
\begin{equation}\label{eq:rayleighmie}
P_{\rm R+M}(\alpha,\lambda) = {P_{\rm Ray}(\alpha,\lambda) + c(\lambda) \cdot P_{\rm Mie}(\alpha,\lambda, \langle r_{\rm eff} \rangle, \langle n_{\rm re} \rangle)}
,\end{equation}
where parameter $c(\lambda)$ ensures the relative weighting of the Mie and Rayleigh components, and may be wavelength dependant. An analogous parametrisation was also used by  \citet{Alexandrov:2012bd} to retrieve cloud properties using polarised reflectance measurements by airborne instruments.

Mie scattering in water droplets in clouds contributes to the polarisation curve in particular at smaller phase angles. For single scattering, the Mie component can  be directly calculated by the scattering matrix elements that depend on the scattering angle, the wavelength, and the properties of the scattering material, in particular its size distribution and its refractory index; see eg, \citet{Hansen:1974en} or \citet{2007AsBio...7..320B}. 

We calculated the scattering phase-matrices using the latest version of {\tt libRadtran (2.0.3)}\footnote{ \url{www.libradtran.org}}, a library of programs to calculate radiative transfer problems in the atmosphere of Earth \citep{Mayer:2005jh, Emde:2016eo}.  Its {\tt mie} utility implements an efficient and well-tested Mie code written by \citet{1980ApOpt..19.1505W} and allows one to calculate the single-scattering phase matrix of spherical particles for a variety of sizes and size distributions. In addition, the optical properties like the (real and imaginary) index of refraction, and their dependence on wavelength can be specified. 

Next, we assemble a grid of Mie models to test if the different scattering properties for populations of particles with different sizes and refractory indices have an observable effect. We aim to retrieve those parameters that fit the observed dependence of polarisation on the phase angle best. This process should allow an estimate of the quality of the retrieval mechanism when compared to those parameters expected and characteristic for Earth. 

Particle radii $r$ follow a standard $\Gamma$ distribution with $n(r) \propto r^\alpha \exp{-\Gamma r}$. According to this distribution, 
effective radii $\langle r_{\rm eff} \rangle$ and their characteristic size distribution widths, $\langle \varv_{\rm eff} \rangle, $ 
are determined for the entire population.
We systematically vary $\langle r_{\rm eff} \rangle$ in the range of 1 and 14 $\mu$m (in steps of 1 $\mu$m), and  probe two typical size distribution widths ($\langle \varv_{\rm eff} \rangle$ equal to 0.1 and 0.01).

We use the well-known index of refraction $\langle n_{\rm re} \rangle$ of water (having a mean $\langle n_{\rm re} \rangle = 1.33$ at a respective temperature of $T=300$K) and consider its imaginary part and wavelength dependence as used in the code of \citet{1980ApOpt..19.1505W}. The properties of the cloud top layers that cause the cloudbow scattering of Earth should therefore be represented by these reference values. In addition, we construct other sets of models with different refractory indices, varying the real part of refraction from 1.24 to 1.44. We leave the imaginary part equal to zero, and assume no variation of the index of refraction  with wavelength for these cases. The different indices of refraction are supposed to bracket a broad range of material and solutions of possible atmospheric constituents in an astrophysical context. For example, liquid methane  has $\langle n_{re} \rangle = 1.24$ at a temperature of $\approx$100K, which is characteristic of Titan \citep{1992RScI...63.2967B}. On the other hand, sulfuric acid has $\langle n_{re} \rangle = 1.44$ for a mixture of 75\% H$_2$SO$_4$ and 25\% H$_2$O, which is characteristic of Venus \citep{1974JAtS...31.1137H}.  

\begin{figure*}
\resizebox{\hsize}{!}{\includegraphics[width=\textwidth, angle=90]{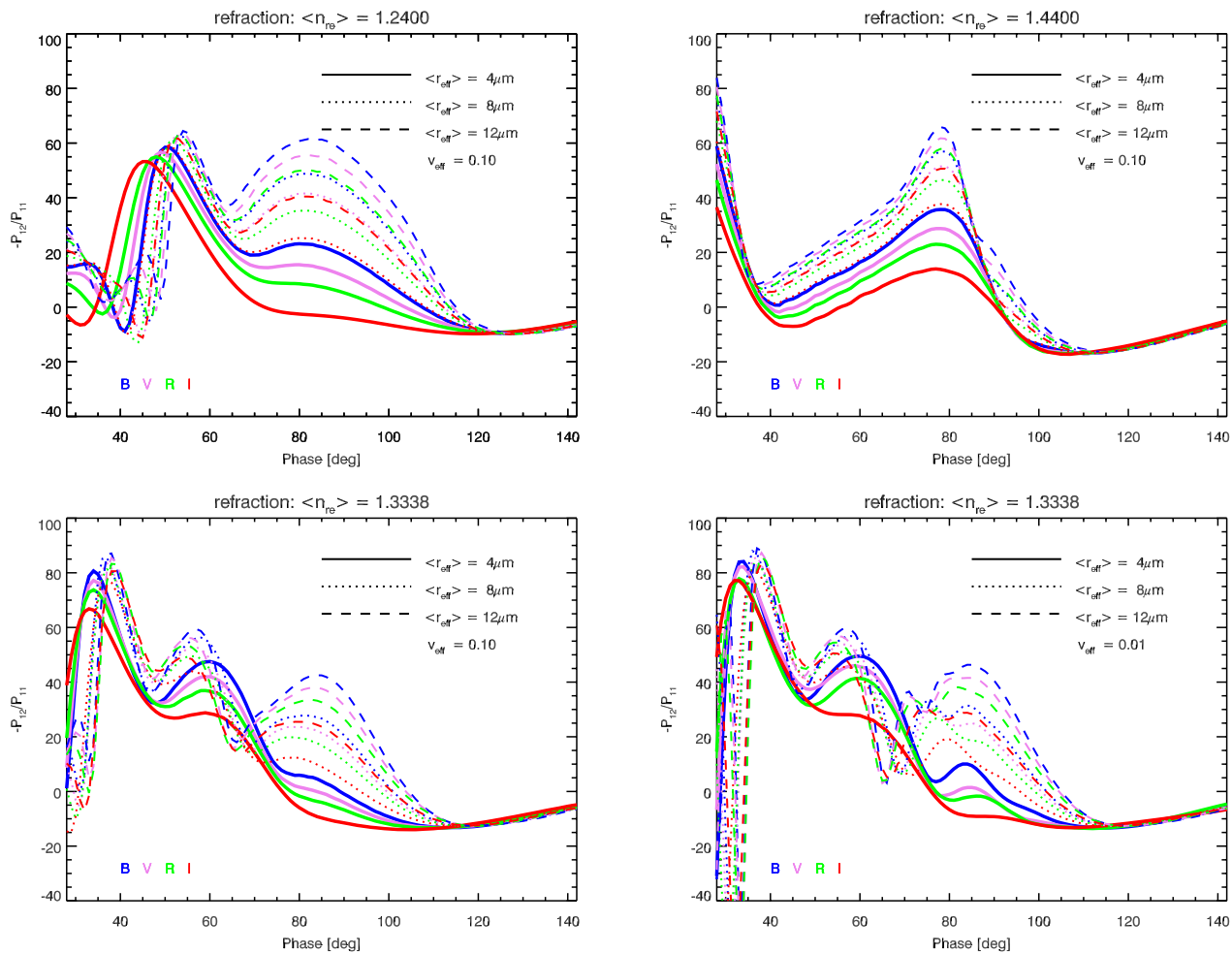}} 
\caption{Examples for calculations of the Mie contribution due to single scattering. The different panels refer to different indices of refraction $\langle n_{\rm re} \rangle$. Each panel plots the ratio of the single scattering phase element $-P_{12}/P_{11}$ as a function of phase angle, for four wavelengths 450nm, 550nm, 650nm, and 850nm ($B, V, R,$ and $I$). Different line styles indicate different mean particle radii distributions for  $\langle r_{\rm eff} \rangle = 4, 8,$ and 12~$\mu$m.    }
\label{Fig:Mie}
\end{figure*}

In Fig.~\ref{Fig:Mie} we show the ratio of the scattering matrix elements $-P_{12}/P_{11}$ as examples of Mie calculations for different indices of refraction, effective widths, and three different mean particle radii. 
As expected, the primary rainbow peak is prominent for all cases, and shifts its peak of maximum polarisation towards smaller phase angles for higher indices of refraction. The secondary rainbow appears at larger phases angles, and is increasingly prominent with increasing mean particle sizes. 
For the refractive index of water, we compare two size distribution widths: 
$\langle \varv_{\rm eff} \rangle = 0.1$ is often considered as a `standard' distribution width of water droplets in Earth clouds. $\langle \varv_{\rm eff} \rangle = 0.01$ is narrower, and the different orders of the single scattering rainbow peaks appear slightly more pronounced. 
We note that $-P_{12}/P_{11}$ is the fractional polarisation in the (unrealistic) case of single scattering only. Nevertheless, we use it equivalent to $P_{\rm Mie}$ in Eq.~\ref{eq:rayleighmie}, heuristically including multiple scattering effects through $dePol$ and $c$.

The grid of Mie-models consists of 17 different choices of refractory indices, each calculated for 14 different mean particle radii from 1 to 14 $\mu$m, and for four wavelengths (450nm, 550nm, 650nm and 850nm, representing the bandpasses derived from the observed spectra in $B,V, R$ and $I$). We use  $\langle \varv_{\rm eff} \rangle = 0.1$ for all values, and in addition $\langle \varv_{\rm eff} \rangle = 0.01$ and $0.02$ for the models with an index of refraction of water.

Final models are constructed according to Eq.~\ref{eq:rayleighmie}. 
For each model, a $\chi^2$ fit is performed in such a way as to minimise the deviation of Eq.~\ref{eq:rayleighmie} to the data simultaneously in all four wavelength bands. 
We consider all measurements of the fractional polarisation obtained so far for the same geometrical viewing aspect of Earth (Pacific) from Paper 1, and listed in Tables~\ref{Tab:Log} and \ref{Tab:PEarth}. All data points together sample the phase curve from 33\degree\ to 138\degree\ and contain information about the Rayleigh {\sl and} the Mie part.   
Parameters $\Delta\alpha$, $W$, $dePol,$ and $c$ in Eq.~\ref{eq:rayleigh} are allowed to vary for each band. Those combinations of parameters that minimise $\chi^2$  are then calculated for each model in the grid. 
In total, we have 144 independent data points (36 phase angles for each of the four wavelength bins). Four parameters are free for each wavelength, which yields a total of 128 degrees of freedom to calculate reduced $\chi^2$ values for each model. 

\begin{figure} 
\resizebox{\hsize}{!}{\includegraphics[trim=70 50 40 50]{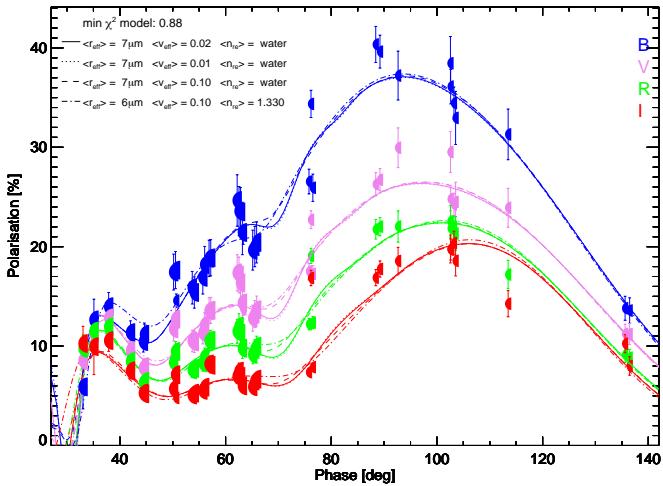}} 
\caption{Fractional polarisation of Earth  as a function of phase angle (Sun--Earth--Moon) with models in B, V, R, and I bandpasses according to Eq.~\eqref{eq:rayleighmie}. The error bars indicate the effects of uncertain lunar albedo on polarisation. The phase curves of four representative models with minimum $\chi^2$ are overplotted with different line styles, together with values of their characteristic sizes and refractory indices.  }
\label{Fig:MRfit}
\end{figure}

Figure~\ref{Fig:MRfit} displays the results of the four best models according to the $\chi^2$-fit procedure, together with the data points of observed fractional polarisation. 
Data points at low phase angles from this work are indicated with larger symbols compared to others from Paper~1.

All four best-fit models appear quite similar, except at very low phase angles below 30\degree, which are not accessible by our Earthshine measurements. 
While the width of the distribution is not constrained by the data, a refractive index of (or around) water and characteristic droplet sizes around 6-7~$\mu$m are retrieved. It is reassuring that models with a very narrow range of indices of refraction around water ($1.33\pm0.01$) are favoured and well constrained given the sensitivity of the rainbow scattering angle on the refractive index.

In addition, the overall shape of the curves represents the measured values quite well for all four bandpasses. The peak value of 15\% (in $B$) and 10\% (in $R$) around the primary rainbow angle is reasonably well retrieved, and the shape of the fitted curves around the rainbow angle is within the uncertainties.  

In Table~\ref{Tab:Phasefits} we summarise the fit parameters for the best-fitting model, which are characterised by the refractive index of water and an effective droplet radius of $r_{\rm eff} = 7\mu$m. 
For this model we estimate the variation of the fit parameters $\Delta\alpha$, $W$, $dePol$ and $c$ that minimise the $\chi^2$ deviations, considering the error bars of the data, which are mainly due to the inaccurate determination of the lunar depolarisation factor and are listed in Table~\ref{Tab:PEarth}.
The upper (and lower) bound are derived from assuming the upper (and lower) polarisation values of the observed data set. 

It is interesting to note that the parameters for $\Delta\alpha$, $W$, $dePol,$  and $c$ typically do not change by more than 10\% among the best-fitting models, which may indicate the robustness of the $\chi^2$ minimisation techniques applied. The parameter $c$ - indicating the Mie contribution to the polarisation -- is only slightly dependant on wavelength, and varies between 13\% (in $I$) and 15\% (in $V$). 

\begin{figure}[t] 
\resizebox{\hsize}{!}{\includegraphics[angle=90,trim=60 70 60 80]{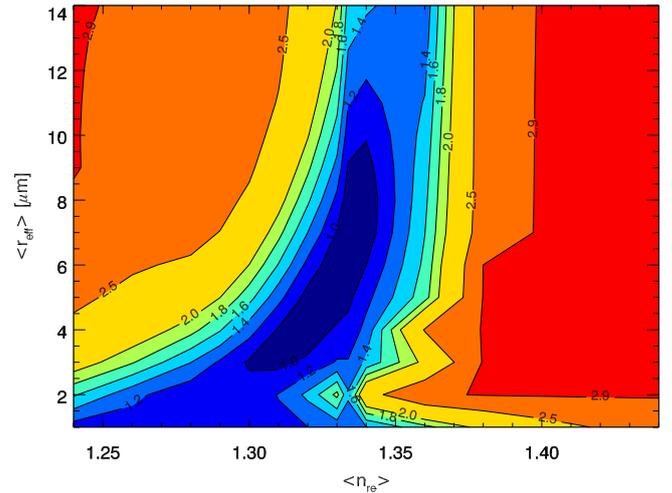}} 
\caption{Contour map of reduced $\chi^2$ to fit the fractional polarisation for sets of models using Eq.~\eqref{eq:rayleighmie}. The regions of minimum $\chi^2 \leq  1$ are coloured in dark blue.  }
\label{Fig:MRfit_X2}
\end{figure}

\begin{table}[h]
\caption{Values of the parameters for  $W$, $\Delta\alpha$, $dePol$ and $c$ for the best model that minimises $\chi^2$ for Eq.~\ref{eq:rayleighmie} shown in Fig.~\ref{Fig:MRfit}. 
All datasets for the Pacific including new ones labelled H.x, I.x, J.x, and K.x in addition to sets B.x, E.x, and G.x from Paper 1 have been used in all four passbands. Errors of the parameters are derived from fits using the error envelope for the data from Table~\ref{Tab:PEarth}.}
\label{Tab:Phasefits} 
\centering
\begin{tabular}{ccccc}
\hline\hline
&  $\Delta\alpha$ [deg] & $W$ & $dePol$ & $c$ \\ 
\hline 
%
$B$ &  8.98$^{+ 0.30}_{- 0.33}$ &  1.65$^{+ 0.07}_{- 0.08}$ &  1.76$^{+ 0.13}_{- 0.15}$  &  0.13$^{+ 0.01}_{- 0.01}$ \\[1mm]
$V$ & 13.41$^{+ 0.50}_{- 0.55}$ &  2.04$^{+ 0.06}_{- 0.07}$ &  2.74$^{+ 0.19}_{- 0.21}$  &  0.15$^{+ 0.01}_{- 0.01}$ \\[1mm]
$R$ & 14.63$^{+ 0.45}_{- 0.50}$ &  2.61$^{+ 0.05}_{- 0.06}$ &  3.27$^{+ 0.23}_{- 0.26}$  &  0.14$^{+ 0.01}_{- 0.01}$ \\[1mm]
$I$ & 17.05$^{+ 0.43}_{- 0.49}$ &  3.16$^{+ 0.06}_{- 0.07}$ &  3.62$^{+ 0.26}_{- 0.29}$  &  0.12$^{+ 0.01}_{- 0.01}$ \\[1mm]
\hline
\hline
\end{tabular}
\end{table}
The goodness of fit quality for all models is displayed in Fig.~\ref{Fig:MRfit_X2}. The contour levels indicate the same regions of reduced $\chi^2$ in the grid of models that vary the index of refraction (x-axis) and the mean effective particle radius (y-axis). The lowest contour levels of one-$\sigma$ are indicated by dark blue, and `banana-shaped' region indicates the best-fitting models according to this procedure. The best-fitting models have a narrow region for the indices of refraction around 1.32 to 1.33 and effective radii from 4 to 8~$\mu$m. 

We conclude that a simple, heuristic, two-component model for the polarisation phase curve according to Eq.~\ref{eq:rayleighmie} describes the global polarisation of Earth rather well. It allows robust retrieval of the refractive index and droplet size of liquid water clouds. 
The most likely material of the scattering agent consistent with our measurements is liquid water with droplet sizes around 6-7~$\mu$m.

\subsection{Comparison with radiative transfer models}\label{mystic}

More realistic models of a planetary atmosphere with opaque clouds and reflecting surfaces must include multiple scattering and radiative transfer to correctly calculate the wavelength-dependant Stokes vectors. A full three-dimensional geometry, inhomogeneous surfaces, molecular absorption, and complex micro-physical cloud properties all pose special challenges for solving the problem for all wavelengths and in all directions, which is required to calculate the phase curve. Several simplifying assumptions are usually made to simulate polarisation spectra of planets in general, and of Earth in particular. 

Models by \citet{Stam:2008ij} allow the linear polarisation spectra of Earth to be simulated for all phase angles and  resolve spectral lines. This latter author approximates Earth's inhomogeneous surface albedo (e.g. that of ocean and clouds) by weighted sums of flux reflected on horizontally homogeneous surfaces. Surfaces are covered by an anisotropic Rayleigh scattering atmosphere which itself contains Mie scattering elements that mimic simple water clouds (of a fixed optical depth of ten). This approach can be used to approximate an arbitrary mixture of different components by building linear combinations of weighted sums of polarised fluxes from each constituent coming from scattering and reflection from horizontally homogeneous surfaces. This method allows  the ratio of clouded to non-clouded surface regions found in Earthshine observations to be constrained in particular, as demonstrated in \citet{Sterzik:2012gk} for phase angles near quadrature. In principle, Stam's models also contain signatures of rainbow (cloudbow) that can be compared with our data. While the models show important qualitative similarities to our data, they were not intended to, and do not quantitatively fit the phase curve of Earth for any realistic linear combination of clear ocean surfaces and homogeneous water clouds.  

The explanation of cloudbows requires a more realistic description of water clouds, which adds considerable complexity to these models. Clouds on Earth are typically a mixture of high-altitude, moderate-optical-depth ice clouds, and low-altitude, large-optical-depth water clouds, each with different and characteristic scattering properties. For example, models of the Earthshine polarisation spectra from \citet{Sterzik:2012gk} require the presence of a thin layer of high-altitude ice clouds which effectively increases the polarisation in the red parts of the spectra, as demonstrated in  \citep{Emde:2017eea} for phase angles around quadrature. 
On the other hand, ice clouds above liquid water clouds are expected to dampen the effect of a rainbow in the polarisation phase curve, but the feature may still be detectable according to the simulations of \citet{Karalidi:2012fc}.  
In addition, clouds on Earth do not appear as homogeneous entities, but are highly structured and geometrically constrained on many scales in horizontal and vertical dimensions. \citet{Trees:2019fc} account for patchy clouds above ocean surfaces for which the effect of a sunglint (specular reflected light on the ocean surface) boosts the polarisation spectra at all wavelengths. 
Their models include water clouds of an optical thickness of 5, with droplets having a standard size distribution ($\langle r_{\rm eff} \rangle = 10\mu$m and $\langle \varv_{\rm eff} \rangle = 0.1$). However, the resulting phase curves overestimate the maximum polarisation at the rainbow angle. Their models predict a maximum polarisation of more than 20\%, which is significantly larger than the percentage that we observe. While their maximum polarisation at the cloudbow peak is relatively independent of the fractional cloud coverage, their polarisation phase curves for angles $\gtrapprox 60\degree$ are more sensitive on the fractional cloud coverage, as expected for its impact on the Rayleigh part at higher phase angles.

As demonstrated with the heuristic model, our observations of cloudbow are sensitive on the index of refraction and the mean effective droplet radii. In addition, multiple scattering in clouds, and thus the average optical thickness of water cloud decks, is expected to have a significant impact on the quantitative agreements of the observed and modelled phase curves.
In the following, we therefore build a grid of models following the approach of \citet{Stam:2008ij} using fractional weighting and linear combinations of different homogeneous surface types. 
Simulations are performed with the spherically symmetric version of the radiative transfer model MYSTIC adapted to Earthshine \citep{Emde:2017eea, Mayer:2009hla, Emde:2010vi}. MYSTIC
is a versatile Monte Carlo code to solve radiative transfer problems in the atmosphere of Earth, and is also available as part of the {\tt libRadtran} package.
The output consists of a polarisation spectrum with all four Stokes parameters ($I, Q, U, V$).

We simulate three generic constituents embedded in or below a mid-latitude summer standard atmosphere. 
As ground surface we assume a reflecting open ocean  with a realistic bi-directional polarisation distribution function \citep{1997JGR...10216989M} including foam caps and shadowing effects by waves (excited by a constant wind speed of 10m/s). As a second component we simulate liquid water clouds  located in an atmosphere at a height between 3 and 4km. We calculate models with different values of optical thickness $\tau_{\rm wc}$ of 5, 7, 10, 15 and 20 in order to analyse the effect of different optical depths of water clouds. Water droplet parameters are assumed as above: spherical droplets follow a $\Gamma$ distribution  with an effective radius $\langle r_{\rm eff} \rangle = 6 \mu$m and a distribution width of $\langle \varv_{\rm eff} \rangle = 0.1$. For comparison we also probe effective radii of $\langle r_{\rm eff} \rangle = $ 5, 7 and 12$\mu$m. 
As a final component we simulate ice clouds located at a height of between 10 and 11km with an optical thickness of one. For the optical properties of the ice cloud,  the parameterisations for a general habit mixture from \citet{Baum:2014gl} were used, with  $\langle r_{\rm eff} \rangle = 30 \mu$m.
Polarisation spectra for each model and for each phase angle between 1 and 179\degree\ were then calculated in steps of 1\degree. We used $10^6$ photons for each Monte Carlo run, typically lasting 30-60 minutes of computing time on a standard PC for a single phase angle, with larger computing times for larger cloud optical thicknesses.  Spectral importance sampling \citep{Emde:2011eg} and the REPTRAN molecular absorption parametrisation \citep{Gasteiger:2014gg}  yield a medium spectral resolution of about 1000 for wavelengths between 4000\AA\ and 10000\AA.  Efficient Monte Carlo simulations including realistic scattering phase matrices of clouds require sophisticated variance reduction methods, which are also included in MYSTIC \citep{Buras:2011hx}. 

We build a grid of models by systematic linear combinations of different fractions $f_x$ for each of the three generic constituents, namely:  water clouds (wc), ice clouds (ic), and open ocean (oo), and denote these fractions with $f_{\rm wc}, f_{\rm ic}$, and $f_{\rm oo}$, respectively. We vary the fractions $f_x$ in steps of 0.1 such that the boundary condition $f_{\rm wc} + f_{\rm ic} + f_{\rm oo} = 1$ is fulfilled. Stokes parameters ($I,U,Q$) for the final model are combined as follows: 
\begin{equation}\label{eq:fractions}
I(Q,U) = f_{\rm wc}\cdot I(Q,U)_{\rm wc} + f_{\rm ic}\cdot I(Q,U)_{\rm ic} + f_{\rm oo}\cdot I(Q,U)_{\rm oo} 
.\end{equation}

The degree of polarisation for the final model is calculated as usual by $P=\sqrt{(Q^2+U^2)}/I$. 
The model grid thus encompasses a total of 66 combinations of the three different components, 
with each scenery for
all 179 phase angles. 
Each model spectrum is then compared to the observed polarisation at a specific phase angle using only data points of the current observations for phase angles between 30\degree\ and 70\degree . For the simulated spectra, we derive the degrees of polarisation in the same four passbands as for the observations to avoid biases. We interpolate the phase angle observed from neighbouring grid points if necessary, and calculate the difference of the measured from the simulated polarisation for each bandpass and phase angle. In this way, $\chi^2$-values are obtained for each model with 84 degrees of freedom  (21 distinct phase points $\times$ 4 bands). 

In Table~\ref{Tab:MC} we list the six best models that minimise the $\chi^2$ values for all models computed  in all four passbands simultaneously. 
\begin{table}[h]
\caption{Characteristics of the six best radiative transfer models that minimise the deviations from the observed polarisation values, sorted by their $\chi^2$ values. }
\label{Tab:MC} 
\centering
\begin{tabular}{cccccc}
\hline\hline
$\chi^2$ &  $\langle r_{\rm eff} \rangle$ & $\tau_{\rm wc}$ & $f_{\rm wc}$ & $f_{\rm ic}$ & $f_{\rm oo}$ \\ 
\hline 
2.12 & 6 & 15 & 0.3 & 0.6 & 0.1 \\
2.14 & 6 & 15 & 0.2 & 0.8 & 0.0 \\
2.17 & 6 & 20 & 0.3 & 0.5 & 0.2 \\
2.22 & 6 & 10 & 0.3 & 0.7 & 0.0 \\
2.23 & 7 & 10 & 0.3 & 0.0 & 0.7 \\
2.23 & 6 & 10 & 0.3 & 0.0 & 0.7 \\
\hline
\hline
\end{tabular}
\end{table}
All good models require $\langle r_{\rm eff} \rangle = 6-7  \mu$m. Smaller or larger $\langle r_{\rm eff} \rangle$ yields a larger $\chi^2$. This result corroborates and confirms the finding of the best-fitting droplet size parameter already obtained in section~\ref{heuristic} with the heuristic model. 

However, the simulations put additional constraints on the optical depth parameter of the water clouds $\tau_{\rm wc}$ and the relative fractions of the generic constituents, $f_{\rm wc}$, $f_{\rm ic}$, and $f_{\rm oo}$. Most importantly, {\sl all} models with $\tau_{\rm wc} < 10$ yield significantly poorer fits. For example, the best-fitting model with  $\tau_{\rm wc} = 7$ already increases $\chi^2 = 2.7$, 
i.e. a quality significantly below
the models with larger $\tau_{\rm wc}$. Even lower $\tau_{\rm wc} = 5$ yields a $\chi^2$ of only $\approx 5$.

\begin{figure}[t] 
\resizebox{\hsize}{!}{\includegraphics[trim=50 50 50 50]{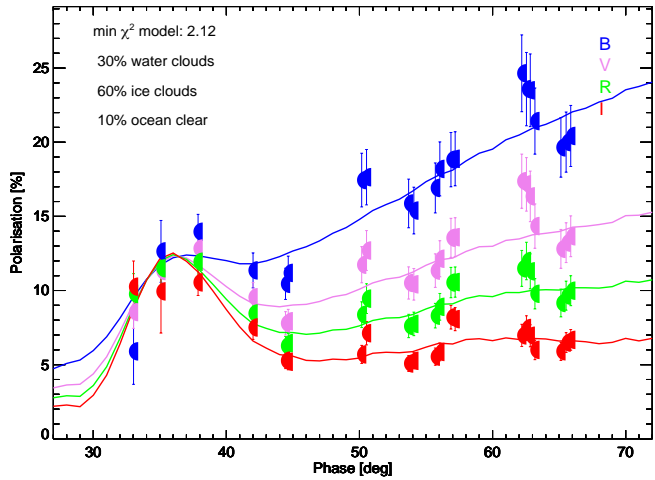}} 
\caption{Fractional polarisation obtained from the best-fitting radiative transfer model simulated with MYSTIC. Surfaces are assembled as homogeneous mixtures of water clouds, ice clouds, and clear ocean, as explained in the text. The best model yields $\langle r_{\rm eff} \rangle = 6 \mu$m,
$\tau_{\rm wc}=15$,
$f_{\rm wc}=0.3$, $f_{\rm ic}=0.6$ and $f_{\rm oo}=0.1$.}
\label{Fig:Earth_Mystic}
\end{figure}

The results of the best-fitting model are displayed in Fig.~\ref{Fig:Earth_Mystic} for all passbands, together with the data. The best model has a fractional contribution of 30\% water clouds, 60\% ice clouds, and 10\% ocean, and fits all four bandpasses with a reduced $\chi^2=2.12$.  We note that the next five best-fitting models (all with $\chi^2 \le 2.3$) are all constrained to fractions of about 30\% water clouds, while the relative fractions of ice clouds and the open ocean surface fractions are much less certain. 
It is interesting to note that the water cloud fraction and its optical depth can be constrained relatively independently. One might expect that, for example, a higher cloud fraction and lower cloud optical thickness would yield a similar degree of polarisation, but it appears that for phase angles around cloudbow, the radiance is only mildly influenced by Rayleigh, ice clouds, and water surface, i.e. $I_{\rm wc} >> I_{\rm Ray} + I_{\rm ic} + I_{\rm oo}$ and $Q_{\rm wc}>>Q_{\rm Ray} + Q_{\rm ic} + Q_{\rm oo}$. In this case, the total polarisation $P$ for the cloudbow is determined by $Q_{\rm wc}/I_{\rm wc}$ only, and does not depend on the individual fractions $f_{\rm i}$. Therefore, for the phase-angle range considered, $P$  only depends on the optical thickness of the cloudy part, allowing its determination.


\section{Discussion}

How do the parameters obtained compare to values that are known or expected for Earth? Only Earth allows us to determine globally averaged values from manifold high-quality resources, such as globally observed and model datasets available for weather and climate modelling and forecasting. On the other hand, our underlying Earthshine polarisation observations are affected by relatively poor phase sampling and considerable measurement errors.  It is therefore not obvious that parameters retrieved match real and/or known values {\sl a priori}. 

The quantitative determination of the correct refraction index of water for Earth might seem trivial. However, the obtained value with a relative accuracy $<$1\% matches the reference value remarkably well, given the very simplified (but robust) heuristic approach used in sect.\ref{heuristic}. This shows that the phase sampling around the cloudbow peak, and in particular the location and reliable observation of the polarisation minimum, is sufficient and allows us to sensitively constrain refraction caused by Mie scattering. 

The quantitative determination of correct mean droplet sizes and the water cloud optical depths is less obvious. 
While their values are known qualitatively, they -- unlike the index of refraction -- vary spatially and temporarily over Earth. Moreover, these parameters are not simple observable quantities for which validated, globally averaged values are available. 

In order to compare our values, we derive mean droplet sizes and cloud optical depths based on ECMWF meteorological data, available as ERA5\footnote{ \url{https://cds.climate.copernicus.eu/cdsapp\#!/dataset/reanalysis-era5-pressure-levels?tab=overview}} hourly data for 37 pressure-altitude
levels on a spatial grid of 0.25\degree\ $\times$ 0.25\degree .
The ECMWF meteorological data offer no direct information on the effective radius of cloud droplets and ice particles, but only contain data fields for liquid and ice water content. The water cloud effective radius $r_{\rm eff}$ can therefore only be expressed indirectly.
It is usually expressed as a function of liquid water content $L$ and droplet number concentration $N_{d}$ using the relation \citep{Martin:1994cr, 2000QJRMS.126.3309W}:
\begin{equation}
r_{\rm eff}=\left(\frac{3E_dL}{4\pi\rho_{w}N_{d}k}\right)^{1/3}
,\end{equation}
where $\rho_{w}$ is the density of liquid water and $k$ is a constant that relates the mean and effective droplet sizes. For water surfaces, $k$ has a value of 0.77. $N_{d}$ depends on the aerosol number concentration $N_{a}$ below the cloud base, which can be calculated as a function of the aerosol mass concentration $q_a$ \citep{Boucher:2017dt, Lowenthal:2004ck}, which itself depends on the surface wind speed $W$ \citep{1986JGR....91.1067E, 1992TellB..44..371G}, and whose zonal and meridional components are ERA5 products. The effective radius enhancement factor $E_d$ \citep{2000QJRMS.126.3309W} is dependant on the ratio of drizzle water content to cloud water content and its value is slightly greater than one on average.
In addition, the fractional cloud cover $f_i$ is available for each vertical layer. Altogether, the optical depth $\tau_i$ and the cloud effective radius $r{_{\rm eff,i}}$ can be calculated per layer $i$ and can be vertically averaged.  The vertically averaged cloud cover $f_{\rm wc}$ can then be reconstructed taking into account only those layers that have optical depths $\tau$ above a certain threshold. For each observing epoch, we calculated $\langle r_{\rm eff} \rangle$, $\tau_{\rm wc}$, and $f_{\rm wc}$. The values for $\tau_{\rm wc}$ and $f_{\rm wc}$ can be calculated taking into account only those layers that have optical depths above a certain threshold. We use two thresholds, $\tau_i > 0.1$ and $\tau_i >2$, the latter being indicative of cloud cover without the (minor) contributions of translucent and thin clouds. Spatially averaged values were calculated for an area roughly encompassing the Pacific ocean.
Results are summarised in Table \ref{Tab:ERA5}. 

\begin{table}[h]
\caption{Spatially averaged values for $\langle r_{\rm eff} \rangle$ (in$\mu$m), $\tau_{\rm wc}$, and $f_{\rm wc}$ for each observing epoch, reconstructed from ECMWF-ERA5 products. Two threshold values are considered, $\tau_i > 0.1$ and $\tau_i >2$. The two last rows contain mean and standard deviation calculated from the values of the time series.}
\label{Tab:ERA5} 
\centering
\begin{tabular}{lrrrrr}
\hline\hline
Date & $\tau_{\rm wc}^{>0.1}$ & $f_{\rm wc}^{>0.1}$ & $\tau_{\rm wc}^{>2}$ & $f_{\rm wc}^{>2}$ &  $\langle r_{\rm eff} \rangle$ \\ 
\hline 
2011-06-08T00:00  & 13.3 & 0.79  &  14.6  & 0.33  &  6.11 \\
2011-06-09T00:00  & 15.2 & 0.84  &  16.4  & 0.38  &  6.36 \\
2011-06-10T00:00  & 15.8 & 0.81  &  17.1  & 0.36  &  6.41 \\
2012-12-17T00:00  & 16.3 & 0.79  &  17.6  & 0.35  &  6.34 \\
2012-12-18T00:00  & 15.8 & 0.79  &  17.3  & 0.34  &  6.33 \\
2012-12-19T00:00  & 14.8 & 0.79  &  16.2  & 0.36  &  6.25 \\
2013-02-18T00:00  & 13.8 & 0.78  &  15.0  & 0.29  &  6.22 \\
2013-02-19T00:00  & 15.3 & 0.78  &  16.5  & 0.33  &  6.14 \\
2013-02-20T00:00  & 15.9 & 0.79  &  17.1  & 0.35  &  6.12 \\
2013-02-22T00:00  & 15.2 & 0.80  &  16.5  & 0.34  &  6.16 \\
2019-10-31T00:00  & 13.9 & 0.81  &  15.1  & 0.33  &  6.32 \\
2019-11-01T00:00  & 14.5 & 0.78  &  15.7  & 0.35  &  6.30 \\
2019-11-02T00:00  & 14.2 & 0.79  &  15.5  & 0.35  &  6.27 \\
2019-11-30T00:00  & 15.1 & 0.78  &  16.5  & 0.35  &  6.46 \\
2019-12-01T00:00  & 16.8 & 0.76  &  18.4  & 0.36  &  6.49 \\
2019-12-02T00:00  & 15.7 & 0.76  &  17.0  & 0.35  &  6.27 \\
2019-12-29T00:00  & 15.9 & 0.74  &  17.5  & 0.31  &  6.43 \\
2019-12-30T00:00  & 15.2 & 0.74  &  16.7  & 0.32  &  6.28 \\
2019-12-31T00:00  & 14.9 & 0.77  &  16.3  & 0.34  &  6.29 \\
2020-01-28T00:00  & 14.2 & 0.77  &  15.5  & 0.29  &  6.21 \\
2020-01-30T00:00  & 14.9 & 0.74  &  16.5  & 0.29  &  6.20 \\
\hline
mean              & 15.1 & 0.78  &  16.4  & 0.34  &  6.28 \\
standard deviation& 0.9  & 0.02  &   0.9  & 0.02  &  0.11 \\
%
\hline
\hline
\end{tabular}
\end{table}

Table~\ref{Tab:ERA5} shows that the values derived from ERA5 data are quite narrowly distributed around their mean values. The optical depths of the liquid clouds are mostly in the range of 14--17, with mean effective droplet radii of $\langle r_{\rm eff} \rangle = 6.28\pm 0.11 \mu$m. Obviously, cloud fractions are different for different threshold values. Considering thicker clouds only in each vertical layer,  $f_{\rm wc}^{>2} = 0.34 \pm 0.02$.
These values compare favorably to those retrieved in Sects.~\ref{heuristic} and \ref{mystic}. 

On the other hand, globally averaged values of water droplet sizes  on top of cloud decks have not been directly observed so far.  Experiments carried out by airborne or satellite instruments are sensitive to more regional effects, and typically retrieve somewhat larger sizes of about 10~$\mu$m 
\citep[see e.g.]{2004ACP.....4.1255M, Breon:2005cd, Polonik:2020ee}. Particularly over sea surfaces, effective radii up to 11~$\mu$m have been derived from ISCCP data \citep{1994JCli....7..465H}, but smaller sizes of 5~$\mu$m have also  recently been been reported \citep{Alexandrov:2018fv}.  Research is ongoing as to how the droplet sizes change at cloud tops due to mixing and evaporation. The impact of changes in droplet size on the global albedo and energy balance of Earth remain to be investigated. 

We note that the  water cloud fraction and the optical depths of water clouds  also appear consistent with the findings of \citet{Stubenrauch:2013hn}, who provide an assessment of global cloud properties derived from various different Earth observation satellites. 

\section{Conclusion}

We observed polarisation spectra of Earthshine at phase angles sampling the characteristics of rainbow and cloudbow scattering features caused by water droplets at the top of liquid clouds.
The cloudbow feature seen above the Pacific ocean is very prominent, reaching a maximum measured polarisation of about 14\% at a phase angle of 37\degree\ in the $B$ band. 

These observations can be explained by Mie scattering from water clouds in Earth's atmosphere. A simple heuristic model with single-scattering Mie cross-sections in combination with a Rayleigh scattering atmosphere explains the degree of polarisation measured in all optical passbands for phases between 30\degree\ and 140\degree. Without {\sl a priori} knowledge of the physical and chemical scattering components, it sensitively constrains the refractive index and the mean droplet sizes. 
The derived values are consistent with the properties of clouds on Earth. 
In particular, the cloud droplet effective radius determined to be around 6-7~$\mu$m is of specific relevance; it is the first direct measurement of water droplet sizes $\langle r_{\rm eff} \rangle$ at cloud tops above large portions averaged over the Pacific ocean. 

In principle, this retrieval method may therefore also be applicable to other planets beyond the Solar System, provided their phase curves can be measured with similar accuracy. However, in these cases, the effects of higher measurement errors and/or a poorer phase coverage must be considered. These latter could for example be modelled using Markov Chain Monte Carlo methods. 

We increased the fidelity of the heuristic method by detailed vector radiative transfer calculations including multiple scattering in a spherically symmetric approximation of a reflecting ocean surface beneath a standard atmosphere with water and ice clouds. The best models allow us to consistently fit the four bandpasses of the polarimetric phase curve over the visible wavelength regime with good quality. We can corroborate the effective radius of water droplets obtained from the simple model.  
Moreover, these simulations allow us to infer cloud fractions and their optical depths, given the good sampling of angles across the Mie- and Rayleigh-scattering-dominated regimes in the phase curve.   
The values retrieved, in particular for water clouds, are consistent with those from \citet{Stubenrauch:2013hn} and those derived from ECMWF-ERA5 models. We find a higher uncertainty in the fractions of ice clouds and ocean. This might decrease when taking into account even larger phase angles which are more sensitive to the sun glint. 

The models discussed are simply an approximation of the real situation of Earth during our observations. We are aware that the patchiness of the cloud spatial distribution and the distribution among different cloud phases have to be taken into account to better approximate the observing scenery
at any given observation epoch. Therefore, we will pursue three-dimensional Monte Carlo calculations to account for the inhomogeneous surface geometry and the variations of actual cloud cover in the near future. 

\begin{acknowledgements}
Based on observations collected at the European Southern Observatory under ESO programme P104.C-0048. Observations were carried out in "delegated visitor mode" and we thank the Paranal Science Operations team for impeccable support. Archival data from ESO programs P87.C-0040 and P90.C-0096 were also used. 
\end{acknowledgements}

\bibliographystyle{aa}
\bibliography{allpapers}

\clearpage
\onecolumn

\setcounter{table}{0}

\tiny
\begin{landscape}
\begin{longtable}{llrlrrrrrrrrrrrrr}
\caption{Record of observations: date and starting time of the observation sequence, airmass, grism, exposure times, S--E--M phase-angle  $\alpha$, and aspect/geometry ("P" = pacific side). $\Phi$ is the angle between the normal of the nominal (geometrical) scattering plane and the celestial north pole. The degree of polarization of Earthshine $P^{\rm ES}$, and the angle of polarization $\phi$ as defined in Eqs. \ref{eq:pol} and \ref{eq:ang} is tabulated for the bandpasses $B, V, R$ and $I$. Statistical errors of $P^{\rm ES}$ (in brackets) are calculated from the null-profiles. They are - typically - small and negligible.}\\
\hline\hline
ID & UT-START & AM & Grism & DIT$\times$N$_{\rm cyc}$ & $\alpha$ & Geo.  & $P_B^{\rm ES}[\%]$ & $P_V^{\rm ES}$[\%] & $P_R^{\rm ES}$[\%] &$ P_I^{\rm ES}$[\%]  &
$\Phi$ & $\phi_B$ & $\phi_V$ &$\phi_R$ &$\phi_I$  \\ [1mm]
\hline 
\endfirsthead\caption{continued.}\\
\hline\hline 
ID & UT-START & AM & Grism & DIT$\times$N$_{\rm cyc}$ & $\alpha$ & Geo.  & $P_B^{\rm ES}[\%]$ & $P_V^{\rm ES}$[\%] & $P_R^{\rm ES}$[\%] &$ P_I^{\rm ES}$[\%]  &
$\Phi$ & $\phi_B$ & $\phi_V$ &$\phi_R$ &$\phi_I$   \\ [1mm]
\hline 
\endhead
\hline
\endfoot
B.1 & 2011-06-08T00:32 & 1.6 & 300V &  100$\times$16 &  76 & P & 13.9(0.06)  &  8.6(0.03)  &  6.9(0.04)  &  5.6(0.14)  & 111.9 & 111.9(0.02)  & 111.6(0.01)  & 111.2(0.01)  & 110.4(0.04)     \\
B.2 & 2011-06-08T23:02 & 1.1 & 300V &   90$\times$16 &  88 & P & 16.3(0.07)  & 10.0(0.03)  &  7.9(0.05)  &  5.6(0.11)  & 112.9 & 112.4(0.02)  & 112.1(0.01)  & 111.8(0.01)  & 111.4(0.03)    \\
B.3 & 2011-06-09T00:30 & 1.3 & 300V &  180$\times$16 &  89 & P & 16.0(0.05)  & 10.1(0.03)  &  7.9(0.03)  &  5.9(0.12)  & 112.9 & 112.5(0.02)  & 112.2(0.01)  & 111.9(0.01)  & 111.9(0.03)     \\
B.4 & 2011-06-10T00:47 & 1.1 & 300V &  180$\times$16 & 102 & P & 14.5(0.05)  &  9.3(0.03)  &  7.8(0.03)  &  6.5(0.17)  & 112.5 & 111.6(0.01)  & 111.1(0.01)  & 110.6(0.01)  & 110.7(0.05)     \\
B.5 & 2011-06-10T01:55 & 1.3 & 300V &  300$\times$16 & 103 & P & 13.8(0.11)  &  9.1(0.04)  &  7.9(0.05)  &  6.7(0.25)  & 112.4 & 111.3(0.03)  & 111.0(0.01)  & 110.3(0.01)  & 110.1(0.07)     \\[1mm]
E.1 & 2012-12-17T00:19 & 2.1 & 300V &   60$\times$16 &  50 & P &  6.0(0.03)  &  3.7(0.02)  &  2.5(0.02)  &  1.6(0.03)  &  68.5 &  67.5(0.01)  &  66.3(0.00)  &  64.8(0.01)  &  61.0(0.01)    \\
E.2 & 2012-12-18T00:04 & 1.5 & 300V &   60$\times$16 &  63 & P &  8.6(0.04)  &  5.2(0.03)  &  3.5(0.03)  &  2.0(0.05)  &  67.0 &  66.1(0.01)  &  65.2(0.01)  &  64.1(0.01)  &  61.2(0.01)     \\
E.4 & 2012-12-19T00:15 & 1.4 & 300V &   60$\times$16 &  75 & P & 10.7(0.03)  &  6.6(0.03)  &  4.4(0.03)  &  2.4(0.05)  &  66.2 &  65.6(0.01)  &  65.0(0.01)  &  64.0(0.01)  &  61.1(0.02)     \\
E.6 & 2012-12-19T01:43 & 2.2 & 300V &   60$\times$12+30$\times$4 & 76 & P & 10.3(0.22) & 6.4(0.03)  & 4.4(0.04)  &  2.6(0.05)  &  66.2 &  65.4(0.06)  &  64.8(0.01)  &  63.7(0.01) & 60.9(0.01) \\[1mm]
G.1 & 2013-02-18T02:03 & 2.4 & 300V &   60$\times$16 &  92 & P & 14.9(0.18)  & 11.2(0.17)  &  7.9(0.24)  &  6.1(0.22)  &  78.4 &  78.8(0.05)  &  78.0(0.05)  &  77.4(0.07)  &  75.9(0.06)   \\
G.2 & 2013-02-19T00:06 & 1.4 & 300V &  120$\times$16 & 102 & P & 15.4(0.09)  & 11.1(0.06)  &  8.1(0.06)  &  6.6(0.10)  &  82.5 &  81.0(0.03)  &  79.3(0.02)  &  78.3(0.02)  &  76.1(0.03)    \\
G.4 & 2013-02-19T02:16 & 2.0 & 300V &  120$\times$16 & 103 & P & 13.3(0.18)  &  9.3(0.06)  &  7.7(0.05)  &  6.2(0.11)  &  82.8 &  82.2(0.05)  &  81.1(0.02)  &  80.2(0.01)  &  79.1(0.03)     \\
G.5 & 2013-02-20T00:26 & 1.4 & 300V &  120$\times$16 & 113 & P & 12.7(0.14)  &  9.1(0.12)  &  6.2(0.12)  &  4.8(0.23)  &  86.9 &  84.1(0.04)  &  80.7(0.04)  &  79.7(0.04)  &  76.8(0.07)     \\
G.7 & 2013-02-22T01:11 & 1.4 & 300V &   90$\times$16 & 135 & P &  5.6(0.05)  &  4.3(0.04)  &  3.3(0.03)  &  3.5(0.06)  &  94.3 &  91.9(0.02)  &  89.2(0.01)  &  87.8(0.01)  &  86.1(0.02)    \\
G.9 & 2013-02-22T03:03 & 1.4 & 300V &   80$\times$16 & 136 & P &  5.6(0.05)  &  4.3(0.04)  &  3.2(0.03)  &  2.7(0.19)  &  94.3 &  93.0(0.02)  &  90.8(0.01)  &  91.3(0.01)  &  89.1(0.06)     \\[1mm]
H.1 & 2019-10-30T23:48 & 2.5 & 300V &   60$\times$8 &  37 & P &  5.8(0.07)  &  5.0(0.04)  &  4.4(0.05)  &  3.6(0.04)  &  93.5 &  93.8(0.02)  &  93.4(0.01)  &  93.5(0.01)  &  93.5(0.01)     \\
H.2 & 2019-10-31T23:32 & 1.6 & 300V &  120$\times$16 &  50 & P &  7.2(0.03)  &  4.5(0.03)  &  3.1(0.02)  &  1.9(0.05)  &  90.0 &  90.3(0.01)  &  90.1(0.01)  &  90.0(0.01)  &  90.5(0.02)     \\
H.3 & 2019-11-01T00:15 & 2.0 & 300V &  120$\times$16 &  50 & P &  7.3(0.16)  &  4.9(0.04)  &  3.5(0.04)  &  2.4(0.08)  &  89.8 &  89.7(0.04)  &  88.7(0.01)  &  88.0(0.01)  &  86.7(0.02)     \\
H.4 & 2019-11-01T23:33 & 1.3 & 300V &  120$\times$16 &  62 & P & 10.2(0.05)  &  6.7(0.04)  &  4.3(0.04)  &  2.4(0.06)  &  85.4 &  85.5(0.01)  &  84.8(0.01)  &  84.6(0.01)  &  84.3(0.02)     \\
H.5 & 2019-11-02T00:16 & 1.5 & 300V &  120$\times$8 &  62 & P &  9.8(0.06)  &  6.7(0.07)  &  4.4(0.06)  &  2.6(0.09)  &  85.2 &  85.8(0.02)  &  85.4(0.02)  &  85.2(0.02)  &  85.1(0.03)     \\
H.6 & 2019-11-02T00:51 & 1.8 & 300V &  120$\times$16 &  62 & P &  9.8(0.04)  &  6.4(0.04)  &  4.2(0.03)  &  2.4(0.06)  &  85.1 &  85.1(0.01)  &  84.4(0.01)  &  84.2(0.01)  &  83.3(0.02)     \\
H.7 & 2019-11-02T01:34 & 2.3 & 300V &  120$\times$8 &  63 & P &  8.9(0.07)  &  5.6(0.06)  &  3.6(0.04)  &  2.1(0.08)  &  85.0 &  85.3(0.02)  &  84.6(0.02)  &  83.9(0.01)  &  82.6(0.02)     \\[1mm]
I.1 & 2019-11-30T00:05 & 2.2 & 300V &   90$\times$8 &  42 & P &  4.7(0.04)  &  3.7(0.02)  &  3.1(0.02)  &  2.6(0.07)  &  82.9 &  83.1(0.01)  &  82.9(0.01)  &  82.4(0.01)  &  83.0(0.02)     \\
I.2 & 2019-12-01T00:07 & 1.6 & 300V &  120$\times$16 &  53 & P &  6.6(0.02)  &  4.1(0.01)  &  2.8(0.02)  &  1.7(0.03)  &  78.5 &  77.8(0.01)  &  77.0(0.00)  &  76.1(0.00)  &  74.3(0.01)     \\
I.3 & 2019-12-01T00:57 & 2.2 & 300V &  100$\times$8+80$\times$8 &  54 & P &  6.4(0.03)  &  4.1(0.02)  &  2.9(0.02)  &  1.8(0.03)  &  78.3 &  76.2(0.01)  &  74.9(0.01)  &  73.2(0.00)  &  69.1(0.01)   \\
I.4 & 2019-12-02T00:19 & 1.4 & 300V &  120$\times$16 &  65 & P &  8.2(0.03)  &  5.0(0.02)  &  3.4(0.02)  &  2.0(0.04)  &  74.3 &  73.6(0.01)  &  72.9(0.01)  &  72.2(0.01)  &  69.7(0.01)     \\
I.5 & 2019-12-02T01:09 & 1.8 & 300V &  120$\times$16 &  65 & P &  8.3(0.03)  &  5.1(0.02)  &  3.6(0.02)  &  2.2(0.04)  &  74.2 &  73.5(0.01)  &  72.9(0.01)  &  72.2(0.01)  &  70.4(0.01)     \\
I.6 & 2019-12-02T01:54 & 2.4 & 300V &   90$\times$8 &  65 & P &  8.5(0.06)  &  5.3(0.03)  &  3.7(0.04)  &  2.3(0.07)  &  74.1 &  74.0(0.02)  &  73.4(0.01)  &  72.4(0.01)  &  70.2(0.02)     \\[1mm]
J.1 & 2019-12-29T00:06 & 2.8 & 300V &   30$\times$2 &  33 & P &  2.4(0.91)  &  3.3(0.27)  &  3.6(0.36)  &  3.5(0.48)  &  78.2 &  66.1(0.26)  &  60.0(0.08)  &  65.9(0.10)  &  66.9(0.14)     \\
J.2 & 2019-12-30T00:17 & 2.1 & 300V &   60$\times$16 &  44 & P &  4.3(0.03)  &  3.0(0.02)  &  2.3(0.02)  &  1.8(0.03)  &  74.4 &  73.6(0.01)  &  72.5(0.01)  &  71.3(0.00)  &  69.0(0.01)     \\
J.3 & 2019-12-30T00:43 & 2.7 & 300V &   30$\times$8 &  44 & P &  4.6(0.06)  &  3.1(0.04)  &  2.4(0.03)  &  1.8(0.05)  &  74.3 &  73.2(0.02)  &  72.1(0.01)  &  70.7(0.01)  &  68.9(0.02)     \\
\textsl{  J.3-1} & 2019-12-30T00:43 & 2.7 & 300V &   30$\times$2 &  44 & P &  4.4(0.06)  &  3.1(0.06)  &  2.6(0.03)  &  2.3(0.05)  &  74.3 &  66.8(0.02)  &  58.4(0.02)  &  56.7(0.01)  &  50.9(0.01)    \\
\textsl{  J.3-2} & 2019-12-30T00:46 & 2.7 & 300V &   30$\times$2 &  44 & P &  5.1(0.09)  &  3.8(0.04)  &  2.8(0.03)  &  2.3(0.05)  &  74.3 &  79.0(0.03)  &  83.6(0.01)  &  83.8(0.01)  &  87.7(0.01)     \\
\textsl{  J.3-3} & 2019-12-30T00:48 & 2.7 & 300V &   30$\times$2 &  44 & P &  5.2(0.06)  &  3.8(0.04)  &  2.8(0.03)  &  2.3(0.04)  &  74.3 &  79.3(0.02)  &  83.7(0.01)  &  84.1(0.01)  &  87.2(0.01)     \\
\textsl{  J.3-4} & 2019-12-30T00:50 & 2.7 & 300V &   30$\times$2 &  44 & P &  4.3(0.07)  &  3.1(0.04)  &  2.6(0.03)  &  2.3(0.04)  &  74.3 &  65.5(0.02)  &  57.4(0.01)  &  56.4(0.01)  &  50.9(0.01)     \\
J.4 & 2019-12-31T00:36 & 1.9 & 300V &  120$\times$8 &  55 & P &  7.0(0.03)  &  4.4(0.02)  &  3.1(0.02)  &  1.9(0.06)  &  71.2 &  70.7(0.01)  &  69.8(0.01)  &  68.6(0.01)  &  66.1(0.02)    \\
J.5 & 2019-12-31T01:21 & 2.6 & 300V &  120$\times$4 &  56 & P &  7.5(0.14)  &  4.7(0.09)  &  3.3(0.08)  &  2.0(0.05)  &  71.1 &  71.0(0.04)  &  70.1(0.03)  &  68.7(0.02)  &  66.2(0.02)    \\[1mm]
K.1 & 2020-01-28T00:05 & 2.9 & 300V &   30+20$\times$3  &  35 & P &  5.2(0.71)  &  4.4(0.20)  &  4.2(0.24)  &  3.4(0.90)  &  73.5 &  85.0(0.20)  &  79.1(0.06)  &  75.1(0.07)  &  78.3(0.26)    \\
K.2 & 2020-01-30T00:17 & 1.9 & 300V &  120$\times$16 &  56 & P &  7.8(0.03)  &  5.2(0.02)  &  3.8(0.02)  &  2.6(0.04)  &  69.8 &  68.7(0.01)  &  68.1(0.01)  &  67.4(0.01)  &  66.0(0.01)    \\
K.3 & 2020-01-30T00:59 & 2.6 & 300V &  120$\times$4 &  57 & P &  7.8(0.12)  &  5.2(0.07)  &  3.8(0.06)  &  2.6(0.06)  &  69.8 &  69.7(0.03)  &  69.1(0.02)  &  68.3(0.02)  &  66.6(0.02)    \\[1mm]
\hline
\hline
\label{Tab:Log} 
\end{longtable}
\end{landscape}

\end{document}